\documentclass[preprint,authoryear,12pt]{elsarticle}
\usepackage{graphicx}
\usepackage{color}

\usepackage{amssymb,bm}
\usepackage{subfigure}
\usepackage{algorithm}
\usepackage{algorithmic}

\usepackage{subfigmat}% matrices of similar subfigures, aka small multiples

\usepackage[%
a4paper=true,%
breaklinks=true,%
colorlinks=true,%
pdfauthor={First Author et al.},%
pdftitle={Template for manuscripts in Advances in Space Research}%
]{hyperref}

\usepackage{bm}

\journal{Advances in Space Research}

\begin{document}

%% Frontmatter
\begin{frontmatter}


% Use the tnoteref command within \title and fnref within \author or \address for footnotes;
% use the corref command within \author for corresponding author footnotes;
% \author{Name\corref{cor1}\fnref{label2}}
% \address{Address\fnref{label3}}

\title{Hybrid Perturbation methods based on Statistical Time Series models}
%\tnotetext[footnote1]{This template can be used for all publications in Advances in Space Research.}

% Use optional labels to link authors explicitly to addresses:

\author[grucaci]{Juan F\'elix San-Juan\corref{cor}}
\ead{juanfelix.sanjuan@unirioja.es}
\cortext[cor]{Corresponding author. Tel.: +34 941299440;  fax: +34 941299460.}

\author[grucaci]{Montserrat San-Mart\'in}
\ead{montse.sanmartin@unirioja.es}

\author[grucaci]{Iv\'an P\'erez}
\ead{ivan.perez@unirioja.es}

\author[grucaci_cibir]{Rosario L\'opez}
\ead{rlgomez@riojasalud.es}

\address[grucaci]{Scientific Computing Group (GRUCACI), University of La Rioja,
26004 Logro\~no, Spain}

\address[grucaci_cibir]{Scientific Computing Group (GRUCACI), Center for Biomedical Research of La Rioja (CIBIR), 26006 Logro\~no, Spain}

\begin{abstract}
In this work we present a new methodology for orbit propagation, the hybrid perturbation theory, based on the combination of an integration method and a prediction technique. The former, which can be a numerical, analytical or semianalytical theory, generates an initial approximation that contains some inaccuracies derived from the fact that, in order to simplify the expressions and subsequent computations, not all the involved forces are taken into account and only low-order terms are considered, not to mention the fact that mathematical models of perturbations not always reproduce physical phenomena with absolute precision. The prediction technique, which can be based on either statistical time series models or computational intelligence methods, is aimed at modelling and reproducing missing dynamics in the previously integrated approximation. This combination results in the precision improvement of conventional numerical, analytical and semianalytical theories for determining the position and velocity of any artificial satellite or space debris object. In order to validate this methodology, we present a family of three hybrid orbit propagators formed by the combination of three different orders of approximation of an analytical theory and a statistical time series model, and analyse their capability to process the effect produced by the flattening of the Earth. The three considered analytical components are the integration of the Kepler problem, a first-order and a second-order analytical theories, whereas the prediction technique is the same in the three cases, namely an additive Holt-Winters method. 
\end{abstract}


\end{frontmatter}

\parindent=0.5 cm

\section{Introduction}

The equations of the perturbed motion of  a satellite   can be written as a set of $3$ second-order or $6$ first-order ordinary differential equations.  The orbit propagation  problem consists in computing the position and velocity of the satellite at a given final time $t_f$, from the position and velocity  at a given initial time $t_1$.  Classically, the techniques used to solve this problem have been  three. 

The first two methods are  known as general and special  perturbation techniques.  General perturbation techniques are based on the analytical integration of the satellite's equations of motion using perturbation theories  \citep{dep1969_canontransf, gia1964_notesvonzei, hor1966_genpertunscanon, hor1971_genpertnoncanon, kry1943_nonlinmech, mor1966_vonzei}. These techniques  provide approximate analytical solutions \citep{aks1970_order2, bro1959_astnodrag, hoo1980_spacetrack3, hoo1987_analdynatmos, kin1977_ord3ast, koz1962_2ordnodrag, lyd1963_smalleccinclbrow} valid for any set of initial conditions. These solutions are explicit functions of time, physical parameters and integration constants, which are mainly characterized by retaining the essential behaviour of the motion. It is  worth noting that most analytical theories currently in use only consider very basic  models of external forces, because in some cases their corresponding analytical expressions can be too  cumbersome. Furthermore, only low-order approximations are taken into account because analytical expansions for the higher-order solutions may become unmanageably long.  Some of these theories can even implement truncated dynamic parameter expansions, so that  their accuracy  and computational efficiency are closely related to the initial assumptions. 

On the other hand, special perturbation methods \citep{ber2004_gaussjackson, kin1989_numint, lon1989_gtdsmaththeor1} refer to  the accurate numerical integration of the equations of motion, including any external forces, even those in which analytical manipulations are complicated, which makes it necessary to use small  steps in order   to  integrate the equations of motion.  General perturbation methods  produce  more computationally efficient propagators although  are not as accurate as those  developed using special perturbation techniques.

Finally, the third approach is the semianalytical technique \citep{cef2010_accessdsst, liu1980_sstclosearth, nee1998_currentdsst}, which combines and  takes advantage of the best characteristics of both the general and special perturbation techniques.  This approach allows  to include any external forces  in the equations of motion, which are simplified  using analytical techniques. Thus,  the transformed equations of motion can be integrated numerically in a more efficient way by using longer integration steps.

Current needs for Space Situational Awareness require improving orbit propagation of space objects in different ways, including the efficient short-term propagation of catalogues of thousands of objects, the accurate very-long-term propagation needed for designing disposal strategies, the instant propagation of fragmentation models, or the propagation of uncertainties in observed orbits of Potentially Hazardous Objects, among others.

Improvement in the models to be integrated constitutes a basic line of research, together with the use of advanced computer architectures based on parallel processing. Additional improvement can be achieved by combining both integrating and forecasting techniques, which we have called hybrid methods.

In this work we present the hybrid perturbation theory, which may combine any kind of the aforementioned integration techniques with forecasting techniques based on statistical time series models \citep{cha2012_tsa_r, tra2015_tseries_r, san2012gru_sarimahop} or computational intelligence methods \citep{per2013gru_nnhop}. This combination allows for an increase in the accuracy of the numerical, analytical or semianalytical theories for predicting the position and velocity of any artificial satellite or space debris object, through the modelling of higher-order terms and other external forces not considered in those initial theories, as well as some physical effects not accurately modelled by the mathematical equations. The final goal of hybrid methodology is to complement the mathematical model of an orbiter dynamics, which is never a completely faithful representation of physical phenomena, with real dynamics provided by real observations, thus yielding a more accurate representation of real behaviour. As a first step in the process to eventually include unmodelled physical effects in the formulation of the problem, we start by considering a basic perturbation, $J_2$, and check the capability of the hybrid propagator to grasp its dynamics. In this process we simulate real observations by means of numerically generated ephemeris through an $8^{th}$ order Runge-Kutta method \citep{dor1989_rungekutta}.

The aim of this paper is to develop a family of hybrid orbit propagators based on three different orders of approximation of an analytical theory,  in order to model the effect produced by the flattening of the Earth so that this technique can be validated. These hybrid orbit propagators incorporate the integration of the Kepler problem in the first case, a first-order analytical theory in the second case and a second-order analytical theory in the last case as the integration techniques; the forecasting technique is an additive Holt-Winters method in the three cases.

This paper is organized as follows. Section 2 describes the concept that underlies hybrid perturbation methodology. Section 3 outlines the second-order analytical theory PPD that, together with its first-order and zero-order approximations, constitutes the base for the three hybrid propagators to be developed in the following sections. Section 4 describes the Holt-Winters method, an exponential smoothing technique used in this paper as the forecasting part of the hybrid propagators. In Section 5, the construction of the three hybrid propagators is detailed, paying special attention to the preliminary statistical analysis of control data, which is important in order to choose the most appropriate sampling rate for the time series to be processed. Results are analysed, and compared to the conventional analytical propagation results, for a set of 9 LEO satellites. Finally, Section 6 summarizes the study and remarks some interesting findings.

\section{Hybrid perturbation methodology}

A hybrid perturbation theory is a methodology for determining an estimation of the position and velocity of any orbiter, which may be an artificial satellite or space debris object, at a final instant $t_f$, in some set of canonical or non-canonical variables, $\hat{\bm{x}}_{t_f}$.

In a first phase, an integration method $\mathcal{I}$ is needed in order to calculate a first approximation, $\bm{x}_{t_f}^{\mathcal{I}}$, from the position and velocity at an initial instant $t_1$, $\bm{x}_{t_1}$: 

\begin{equation}\label{solana}
\bm{x}_{t_f}^{\mathcal{I}} = \mathcal{I}(t_f,\bm{x}_{t_1}).
\end{equation}

This approximation can include  some inaccuracies derived from the facts that, for the sake of manageability of the resulting expressions and affordability of the subsequent computations, not all the external forces are usually taken into account in the physical model, and only low-order approximations are considered. Additional imprecision arises from the fact that mathematical models of perturbations not always depict real physical phenomena with high fidelity.

The error of this approximation for any instant $t_i$, $\mathcal{\bm\varepsilon}_{t_i} $, can be determined if the exact position and velocity $\bm{x}_{t_i}$ is known, usually through a precise observation: 

\begin{equation}
\mathcal{\bm\varepsilon}_{t_i} = \bm{x}_{t_i} - \bm{x}_{t_i} ^{\mathcal{I}}. \label{error} 
\end{equation}

The second phase of the method requires the knowledge of $\bm{x}_{t_i}$  for  $ t_1,\ldots,t_T$, with $t_T <t_f$, in order to build the time series of errors 
$\mathcal{\bm\varepsilon}_{t_1} \ldots \mathcal{\bm\varepsilon}_{t_T} $ that contains the dynamics not present in the approximation generated during the first phase. 
The time elapsed between $ t_1$ and $t_T$ is defined as control period,  $\mathcal{\bm\varepsilon}_{t_1} \ldots \mathcal{\bm\varepsilon}_{t_T} $  as control data and $T$ as the number of  points in the control period. Then the goal is the modelling of such dynamics in order to be able to reproduce it; this task is accomplished by means of statistical techniques in time series analysis or computational intelligence methods. Once it has been done, an estimation of the error at the final instant $t_f$, $\hat{\mathcal{\bm\varepsilon}}_{t_f}$, can be calculated, and consequently the desired value of $\hat{\bm{x}}_{t_f}$ can be determined as:

\begin{equation} \label{forecast}
\hat{\bm{x}}_{t_f}= \bm{x}_{t_f}^{\mathcal{I}} + \hat{\mathcal{\bm\varepsilon}}_{t_f}.
\end{equation}

\begin{figure}[!!htp]
\centering
\includegraphics[scale = .8]{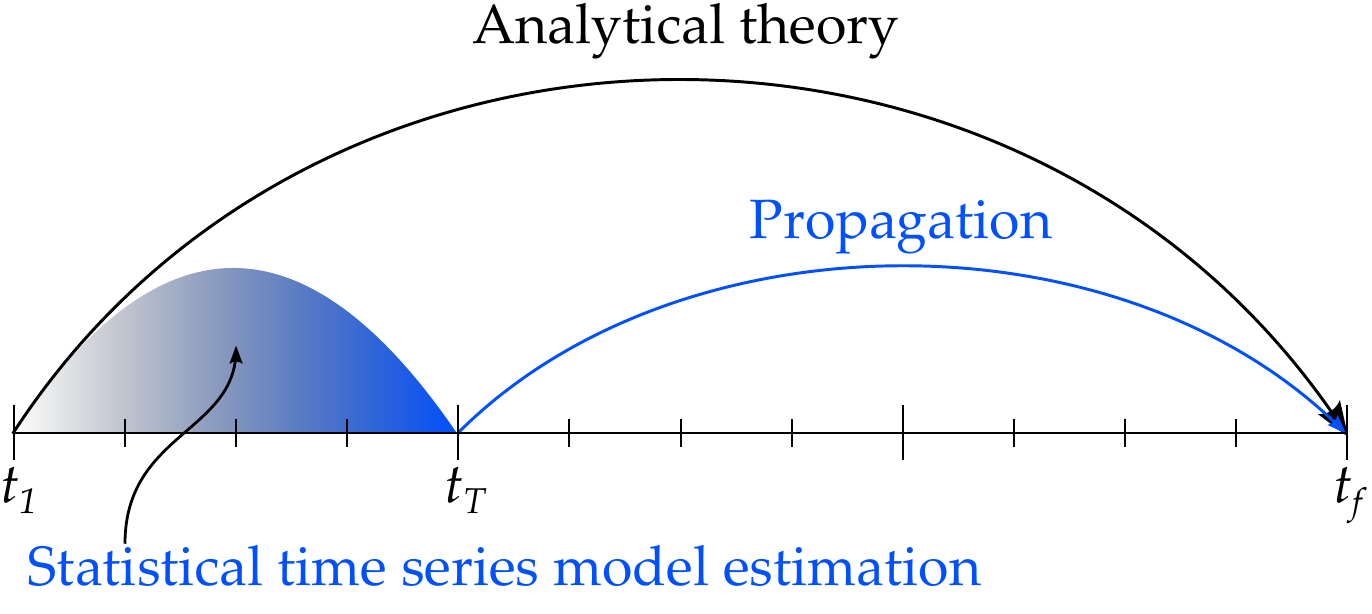}
\caption{Evaluation of a hybrid propagator based on the combination of an analytical theory and a statistical time series model. }\label{fig1}
\end{figure}

It is worth noting that this methodology can be applied to any kind of integration methods regardless of the fact that in this work it has been applied to an analytical theory. For this case,
Figure \ref{fig1} shows the instants at which both the analytical expression and the statistical time series model have to be evaluated.

\section{Second-order analytical theory PPD}\label{astPPD}

The  \textit{main problem} of the artificial satellite theory is defined as a Kepler problem perturbed by Earth's oblateness. This model provides a first approximation to describe the motion of a low Earth orbiter. In this Section we develop a second-order closed-form analytical theory, based on Lie transforms, for the main problem, which we will use to derive the analytical part of the three hybrid orbit propagators to be analysed in Section 5. In the first case, a zero-order approximation, i.e. the Kepler solution, will be considered, whereas a first-order and the complete second-order approximations will be taken into account for the remaining two cases.

The main-problem dynamical system can be described in a cartesian coordinate system  ($\bm{x}$, $\bm{X}$) by means of the Hamiltonian

\begin{equation}\label{ham}
\mathcal{H} = \frac{1}{2} (\bm{X} \cdot \bm{X})  - \frac{\mu}{r} \left[1-J_2 \left(\frac{\alpha}{r} \right)^2 P_2\left(\frac{z}{r}\right)\right],
\end{equation}

\noindent
where  $r=\sqrt{x^2+y^2+z^2}$ is the geocentric distance, $P_2$ represents the Legendre polynomial of degree $2$, $\mu$ is the gravitational constant, $\alpha$ is  the  equatorial radius of the Earth, and $J_{2}>0$ is a constant representing the  shape of the Earth. 

In order to carry out a second-order analytical theory, the Hamiltonian (\ref{ham}) is rewritten in terms of Delaunay variables  $(l,g,h,L,G,H)$. This  set of canonical action-angle variables can  be directly related to the orbital elements through the following expressions:

\begin{equation} \label{relacion}
\begin{array}{lll}
l=M, &\qquad & L=\sqrt{\mu a},\\
g=\omega, &\qquad & G=\sqrt{\mu a (1 -e^2)},\\
h=\Omega, &\qquad & H=\sqrt{\mu a (1 -e^2)} \cos i,
\end{array}
\end{equation}

\noindent
where $M$ is the mean anomaly, $\omega$ the argument of the perigee, $\Omega$ the longitude of the ascending node, $a$ the semi-major axis, $e$ the eccentricity and $i$ the inclination. Therefore,  the transformed Hamiltonian in Delaunay variables yields

\begin{equation} \label{mainpolarnodal}
{\cal H}= -\frac{\mu^2}{2 L^2}  -
 \frac{\epsilon}{2}\frac{\mu}{r}\left(\frac{\alpha}{r}\right)^2(1-3 s^2 \sin^2 (f+g)),
\end{equation}

\noindent
where $\epsilon =J_2$ is  a small parameter, $s=\sin i$ and  $f$ is the true anomaly.

Next, following the  method  described in \citep{dep1969_canontransf}, three Lie transforms are applied in order to remove  the long-period terms due to the argument of the perigee in the first place  \citep{alf1984_perigee, dep1981_parallax}, and then the short-period terms due to the mean anomaly  \citep{dep1982_delaunaynorm}. Finally,  the transformed Hamiltonian  yields, up to second order,

\begin{eqnarray}
\mathcal{K} & = &  -\frac{\mu^2}{2 L''^2} + \epsilon \frac{\alpha^2 \mu^4}{4 \eta''^3 L''^6} (3 s''^2 - 2) 
- \epsilon^2 \frac{3 \alpha ^4 \mu ^6 }{128  \eta''^7 L''^{10}} \left[ \left(5 \eta''^2+36 \eta'' +35\right) s''^4 \right. \nonumber  \\[1ex] && \left.
+8 \left(\eta''^2-6 \eta'' -10\right) s''^2-8 \left(\eta''^2-2 \eta'' -5\right)\right],
\end{eqnarray}

\noindent
where $\eta'' = \sqrt{1-e''^2}$. It is worth noting that $\mathcal{K}$ is independent of the variables $l''$, $g''$ and $h''$, and thus Hamilton's equations can be easily integrated by quadratures.
 
The algebraic manipulations required to develop this analytical theory and its corresponding analytical orbit propagator program were built using a set of \textit{Mathematica} packages called MathATESAT \citep{san2011gru_mathatesat}, which is a reimplementation of the ATESAT \citep{san1994gru_atesat_cnes, san1998gru_atesatvspsimu_cnes}. The acronym PPD  makes reference to the sequence of Lie transforms used to carry out this analytical theory;  in this case, the involved transforms  are the elimination of the Parallax, the elimination of the Perigee and the  Delaunay normalization.

\section{Time series forecasting using  exponential smoothing methods}

Exponential smoothing methods are forecasting algorithms for time series. Their main advantages are their ease of application, speed and reduced computational burden. Predictions generated by these methods are based on previously collected data, giving higher importance to more-recent observations. These methods assume a time series is the combination of three components: the trend or long-term variation, the seasonal component, which represents periodic oscillations that repeat at constant intervals, and the irregular or non-predictable component. There are two main procedures for combining these components, depending on the cyclic behaviour with respect to the trend: the additive and the multiplicative compositions. In the additive case, the series shows stable cyclic fluctuations, independently of the increase in the series level. On the contrary, the multiplicative model implies a change in the amplitude of the seasonal oscillations as the series trend varies. It is worth noting that a multiplicative model can be converted into an additive one through a Box-Cox transformation.  

In mathematical terms, a time series, $\varepsilon_t$, can be decomposed into trend, $\mu_t$, seasonal variation, $s_t$, and irregular or non-predictable component, $\nu_t$. In an additive model, these components combine in the following manner:

\begin{equation}
\varepsilon_t=\mu_t + s_t + \nu_t.
\end{equation}

In particular, the Holt-Winters method  \citep{win1960_forecexpma}  combines a linear trend together with a periodic behaviour. In this method, the trend can be expressed as:

\begin{equation}
\mu_t=a+b t,
\end{equation}

\noindent
where $a$ and $b$ represent the level and slope of the series, respectively.

This method predicts the series value at time t according to the following recursive procedure

 \begin{equation} \label{SE1} 
  \hat{\varepsilon}_{t}=A_{t-1}+ B_{t-1}+S_{t-s},
\end{equation}

\noindent
that is, the addition of level, $A_{t-1}$, and slope, $B_{t-1}$, at the previous instant, plus seasonal variation, $S_{t-s}$, $s$ epochs before, thus being $s$ the period of such seasonal variation.

The corresponding algorithm updates, at every epoch, the level, the slope of the trend and the values of the seasonal factors, by means of three equations. The first equation determines the series level at epoch $t$, $A_t$, as the weighted average of the deseasonalized series value at the same instant $t$ and the non-seasonal prediction at the previous epoch, that is,

\begin{equation} \label{SE2} 
	A_t=\alpha (\varepsilon_t-S_{t-s}) + (1-\alpha) (A_{t-1}+B_{t-1}),
\end{equation}

\noindent
where $\alpha$ is a constant, named smoothing parameter, with values in the interval $[0,1]$.

With the equation

\begin{equation} \label{SE3} 
	B_t=\beta(A_t-A_{t-1}) + (1-\beta) B_{t-1}
\end{equation}

\noindent
the slope can be estimated as the weighted average of the slope at the previous epoch and its corresponding level change. The smoothing parameter $\beta$ can have values in the interval $[0,1]$.

The last equation determines the seasonal component at epoch $t$, $S_t$, as the weighted average of the detrended series and the seasonal value at the equivalent epoch in the previous period,

\begin{equation} \label{SE4} 
	S_t=\gamma(\varepsilon_t-A_{t}) + (1-\gamma) S_{t-s},
\end{equation}

\noindent
where $\gamma$ is another smoothing parameter which can also have values in the interval $[0,1]$.

The smoothing parameters $\alpha$, $\beta$ and $\gamma$ are decisive in the estimation process.  Parameter $\alpha$ controls the smoothing of the level equation, so that low values give more importance to historical data, whereas high values weight recent observations. Parameter $\beta$ modifies the slope estimation in such a way that a value close to $0$ gives more importance to trend, whereas a value near $1$ weights level changes. Finally, $\gamma$ controls the smoothing of the seasonal component, so that high values lead to predictions more sensitive to the series variations.

Algorithm 1  implements the  Holt-Winters method; its inputs are the number of data per revolution,  $s$, the number of revolutions for which precise observations are available, $c$, the epoch number, starting after the last available precise observation, for which the series value has to be predicted, $h$, and the error series  $\{\varepsilon_t\}_{t=1}^{T}$ values with $T=s \times c$. The algorithm is designed to produce $\hat{\varepsilon}_{T+h|T}$ as the output, which represents the time-series forecast at the final instant $t_f=T+h$, based on the time-series value at the end of the control period $T$.

\begin{algorithm}
\begin{algorithmic}[1]
\REQUIRE $s$, $c$, $h$ and $\{\varepsilon_t\}_{t=1}^{T}$
\ENSURE $\hat{\varepsilon}_{T+h|T}$
\STATE Estimate the values of $A_0, B_0,S_{-s+1},\ldots, S_{-1},S_0$
\FOR {$t=1;\,t\leq T;\,t=t+1$}
\STATE $A_t = \alpha (\varepsilon_t-S_{t-s}) + (1-\alpha)(A_{t-1}+ B_{t-1})$
\STATE $B_t = \beta(A_t-A_{t-1}) + (1-\beta) B_{t-1}$ 
\STATE $S_t = \gamma(\varepsilon_t-A_{t}) + (1-\gamma) S_{t-s}$
\STATE $\hat{\varepsilon}_{t} = A_{t-1}+ B_{t-1} + S_{t-s}$
\ENDFOR
\STATE Select \texttt{error$\_$measure} $\in$ \{MSE,MAE,MAPE\} and obtain it as a function of the smoothing parameters
\STATE Obtain the smoothing parameters that minimize   \texttt{error$\_$measure}  using the L-BFGS-B method
\STATE Calculate $A_T, B_T, S_{T-s+1},\ldots, S_{T-1},S_T$  for the optimal smoothing parameters
\STATE  $\hat{\varepsilon}_{T+h|T} = A_T + h B_T + S_{T-s+1+h\,\mathrm{mod}\, s}$
\RETURN  $\hat{\varepsilon}_{T+h|T}$
\end{algorithmic}
\caption{ Holt-Winters }\label{alg1}
\end{algorithm}

The first step consists in estimating the initial values $A_0$, $B_0$, $S_{-s+1},\ldots$, $S_{-1}$ and $S_0$ by means of a heuristic method in which, in the first place, a classical additive decomposition into trend and seasonal variation over the two first revolutions of the satellite is performed. By doing so, the initial values of the seasonal component, $S_{-s+1},\ldots$, $S_{-1}$ and $S_0$, are obtained, whereas the linear regression coefficients over the trend lead to the initial values of level and slope, $A_0$ and $B_0$. Once these values have been obtained, the next step can be undertaken, in order to apply the recursive equations which allow for the calculation of the values of the components $A_t$, $B_t$ and $S_t$, as well as the single-step error prediction $\hat{\varepsilon}_{t}$ for control data, i.e. for $t=1,\ldots,T$ (lines 2--7). These values will remain as functions of the smoothing parameters $\alpha$, $\beta$ and $\gamma$. In the next step, one of the error functions is selected

\begin{equation} \label{med.err}
\begin{array}{l}
  \displaystyle \mbox{MSE}=\frac{1}{T}\sum_{i=1}^{T}{(\varepsilon_t-\hat{\varepsilon}_{t})^2},\\[3ex]
	\displaystyle \mbox{MAE}=\frac{1}{T}\sum_{i=1}^{T}{|\varepsilon_t-\hat{\varepsilon}_{t}|} ,\\[3ex]
	\displaystyle \mbox{MAPE}=\frac{1}{T} \sum_{i=1}^{T}{ \left| \frac{\varepsilon_t-\hat{\varepsilon}_{t}}{\varepsilon_t} \right|} 100,	
\end{array} 
\end{equation}

\noindent
and its value is determined as a function of the smoothing parameters. Next, the values of the smoothing parameters that minimize the chosen error function have to be determined. As it is not easy to minimize the error functions (14) analitycally, a numerical optimization method is necessary. The limited memory algorithm L-BFGS-B, which is one of the most usual ones, has been chosen for that purpose. The L-BFGS-B method \citep{byr1995_limmemopt}, which is a variation of the BFGS method \citep{bro1970_convminalg, fle1970_varmetric, gol1970_varmetric, sha1970_quasinewt}, named after its creators Broyden, Fletcher, Goldfarb and Shanno, is a quasi-Newton limited memory algorithm that allows optimization with restrictions, thus permitting to impose limitations on smoothing parameters.

With the optimal smoothing parameters, the level and slope values for the last control data ($A_T$ and $B_T$) are calculated, as well as the seasonal component values for the last revolution in control data ($S_{T-s+1},\ldots, S_{T-1},S_T$). With these data, it is possible to predict the  value of the series $h$ epochs ahead, $\hat{\varepsilon}_{T+h|T} $ (line 11).

\section{Validation of the methodology}

The proposed  methodology is applied to the \textit{main problem} of the artificial satellite theory so as to model the effect produced by the flattening of the Earth, which corresponds to the $J_2$ term of  Earth's gravitational potential.  Three analytical orbit propagator programs (AOPP)  are  used to conduct this study. They are derived from a  second-order closed-form analytical theory based on Lie transforms, which has  been briefly described in Section \ref{astPPD}. The first AOPP is PPD0, a propagator  derived from  the zero-order analytical theory, in which only the part corresponding to  Kepler's problem has been taken into account. PPD1 is the second propagator; it implements the first-order analytical theory, i.e. the first-order $J_2$ approximation. Finally, PPD2 implements the  second-order analytical theory, i.e. the full  second-order $J_2$ approximation.  From each of these AOPPs, a hybrid analytical orbit propagator program  (HAOPP) has been developed. In these HAOPPs,  statistical time series analysis has been applied to forecast the effects not taken into account in their corresponding initial AOPPs. 

The propagator HPPD0  will be used to demonstrate the capability of this methodology  to model the full $J_2$ effect. It is worth noting that this perturbation is not included in the initial propagator PPD0 at all.  On the other hand, the propagators HPPD1 and HPPD2  will be used to explore   the capability  of this methodology  to model the error introduced by the analytical approximations, $\mathcal{O}(J_2^{2})$ and  $\mathcal{O}(J_2^{3})$ respectively.  In this work,  the additive Holt-Winters method will be used for forecasting  the effects not taken into account in the initial AOPPs.

Finally, in order  to compare and contrast the performance of the HAOPPs,   several tests with numerically-simulated initial conditions corresponding to LEO orbits will be performed.  The  error measure to be considered  will be the distance error over a prediction horizon of 30 days.

\subsection{Data preprocessing}

This methodology starts by choosing the set of variables that will be used for modelling purposes in the forecasting part of the hybrid propagator. In this work,   Delaunay variables have been chosen, although other sets of variables can also be used. After that, in order to build the new propagator, two sets of values corresponding to the same satellite  are necessary during the control period. The first consists of accurate  values, obtained through the numerical integration of the original problem (\ref{ham}) by using a high-order Runge-Kutta method  \citep{dor1989_rungekutta}, which are considered as actual values from precise observations.  The second is obtained by applying the initial integrating part of the hybrid propagator; it contains  approximate values which do not include, either in whole or in part, the effect that we want to model in the forecasting part.  It is worth noting that the control data  should include an amount of values which is enough to identify any pattern that we expect the forecasting part to model and reproduce.

Then,  subtracting both  data sets for each variable the error  time series $(\varepsilon_t^l,\varepsilon_t^g,\varepsilon_t^h,\varepsilon_t^L,\varepsilon_t^G,\varepsilon_t^H)$ are obtained.  After this operation, the angular-variable time series $\varepsilon_t^l$, $\varepsilon_t^g$ and $\varepsilon_t^h$  may include some outliers that differ from the rest of  values in a quantity multiple of $2\pi$. Such differences correspond to complete spins and, although they have no effect on trigonometrical calculations, for  time series analysis they represent abrupt discontinuities in values that are actually very close. Next,  by adding or subtracting complete spins ($2\pi$),  their values  can be  homogenized to the interval $(-\pi,\pi]$, thus avoiding this and other problems  related to the periodic behaviour of these series. 

The time series $\varepsilon_t^H$ is always 0 for the problem considered here, which means that the pure analytical theory is able to determine  $H$ values accurately. Therefore, forecasting of  this time series is not necessary. For each of the remaining time series,  an univariate Holt-Winters model will be developed  from a preliminary analysis  so as to forecast its future values. This analysis includes the study of the sequence graphics, periodograms and autocorrelation functions (ACF). These graphics  can help  reveal the most important  characteristics of the series, such as trend, stationarity, atypical values, etc.  

Table \ref{table1} shows the orbital elements for nine fictitious  satellites used for testing the HAOPPs. These initial conditions correspond to LEO orbits and all of them have been chosen to avoid the intrinsic singularities present in Delaunay variables.

\begin{table}[htbp!!]
\caption{Satellites used for validation studies of hybrid methodology.}\label{table1}
\begin{center}
\begin{tabular}{rcccc}
Id & $a$ (km) & $e$ &  Inclination (deg)  & Period (min)\\ \hline
1 & 7228 & $0.0631$ & $ \:49$ & $101.926$\\
2 &  7872 & $0.1380$ &  $144$ & $115.936$\\
3 &  7612 & $0.1132$ &  $102$ & $110.156$\\
4 &  7674 & $0.1124$ &  $ \:68$ & $111.505$\\
5 &  7064 & $0.0323$ &  $ \:62$ & $\:98.477$\\
6 &  7087 & $0.0504$ &  $ \:73$ & $\:98.958$\\
7 &  6992 & $0.0268$ &  $ \:29$ & $\:96.975$\\
8 &  7269 & $0.0713$ &  $ \:66$ & $102.795$\\
9 &  7128 & $0.0499$ &  $ \:66$ & $99.818$\\
\end{tabular}
\end{center}
\label{table1}
\end{table}%

\subsection{Modelling the full $J_2$ effect}

Satellite 1  will be used in this subsection to  illustrate this methodology.
It is necessary to start mentioning that in the case of  the Kepler problem, the orbital elements are constant over time (with the unique exception of the mean anomaly $M$,  which varies between $0$ and $2\pi$ during each revolution of the satellite). However, adding the $J_2$ effect to the Kepler problem, i.e. the main problem, produces  significant effects on the orbital elements, which include secular, long and short period variations.

Table \ref{table2} shows the distance error between PPD0 and the numerical integration of the main problem at different  instants. 
The distance error grows up to a maximum value of about $14500$ kilometers, which represents approximately the distance between perigee and apogee. These values will be used to evaluate the improvement introduced by the hybrid propagators.

\begin{table}[htbp!!]
\caption{ Maximum distance  error between PPD0 and the numerical integration of the main problem for  satellite 1 over different propagation spans.}\label{table2}
\begin{center}
\begin{tabular}{cc}
Time &   \multicolumn{1}{c}{Distance error (km) }\\ \hline%\\[-2ex]
$17$ hours & $\:\:864.80$  \\
  $\:1$ day & $\:1202.24$  \\
 $\:2$ days & $\:2408.31$ \\
 $\:7$ days & $\:7894.81$ \\
 $30$ days & $14506.50$ 
 \end{tabular}
\end{center}
\label{default}
\end{table}

\begin{figure}[!!htp]
\centering
\includegraphics[scale=.5]{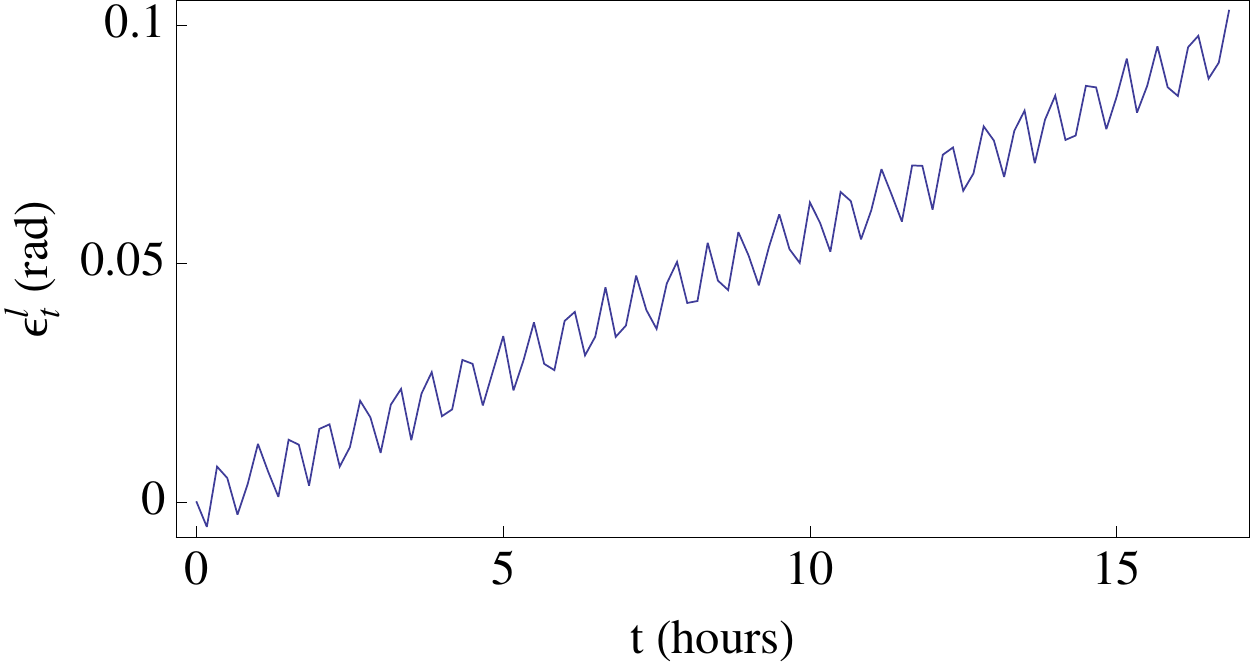}

\includegraphics[scale=.5]{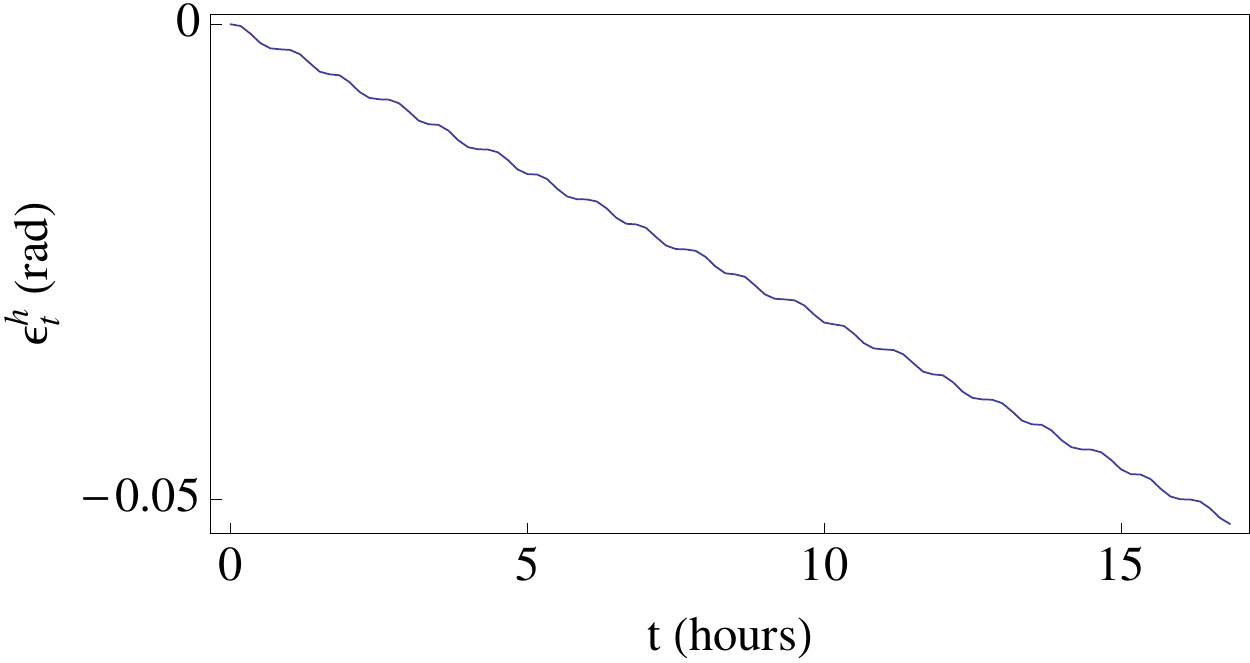}

\includegraphics[scale=.5]{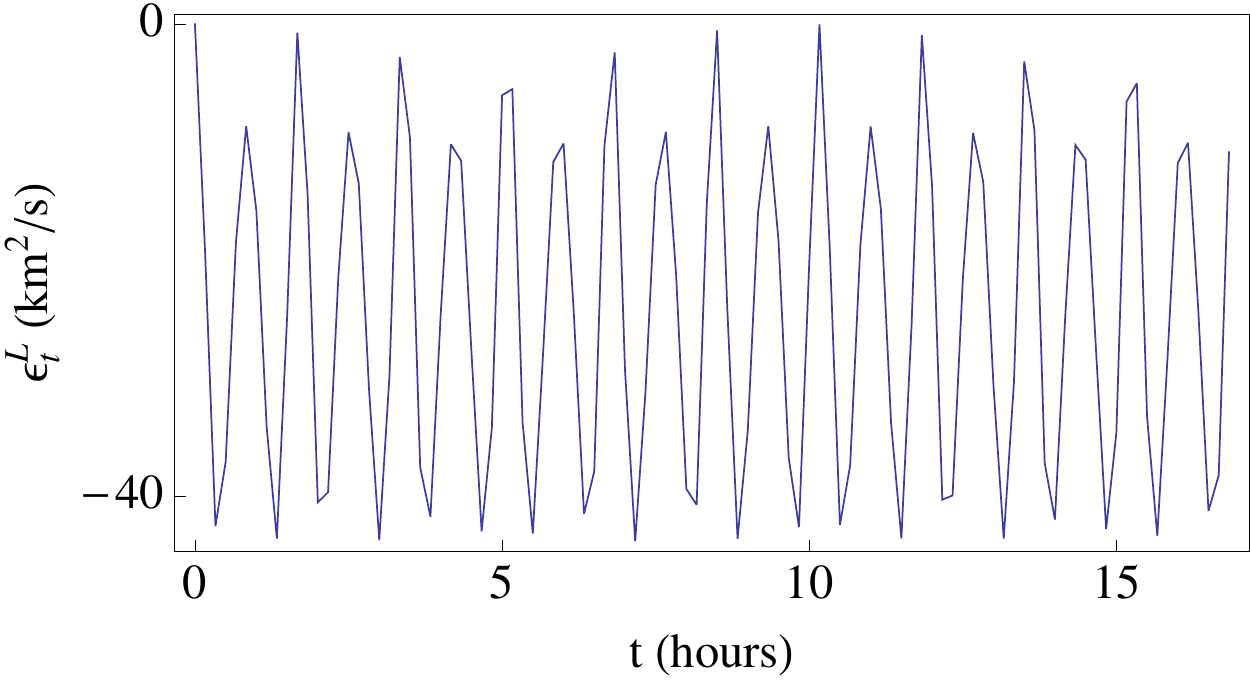}

\caption{Sequence plots of the series $\varepsilon_t^l$, $\varepsilon_t^h$ and $\varepsilon_t^L$.  }\label{figure2a}
\end{figure}

\begin{figure}[!!htp]
\centering

\includegraphics[scale=.5]{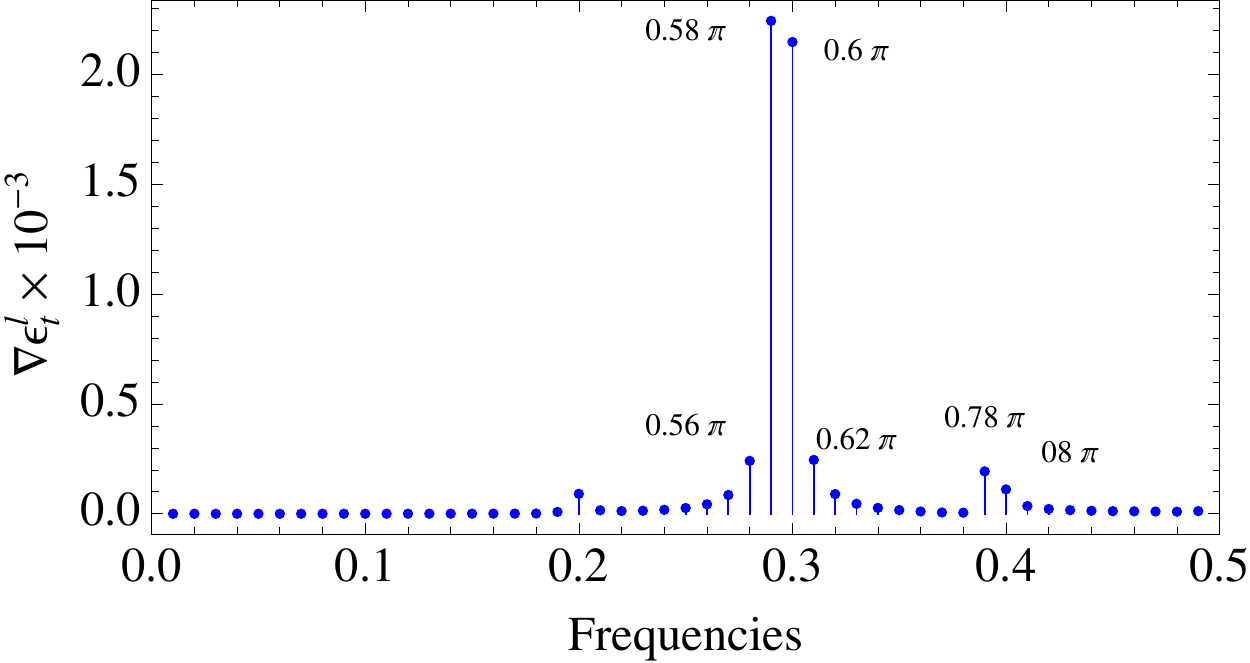}
\includegraphics[scale=.5]{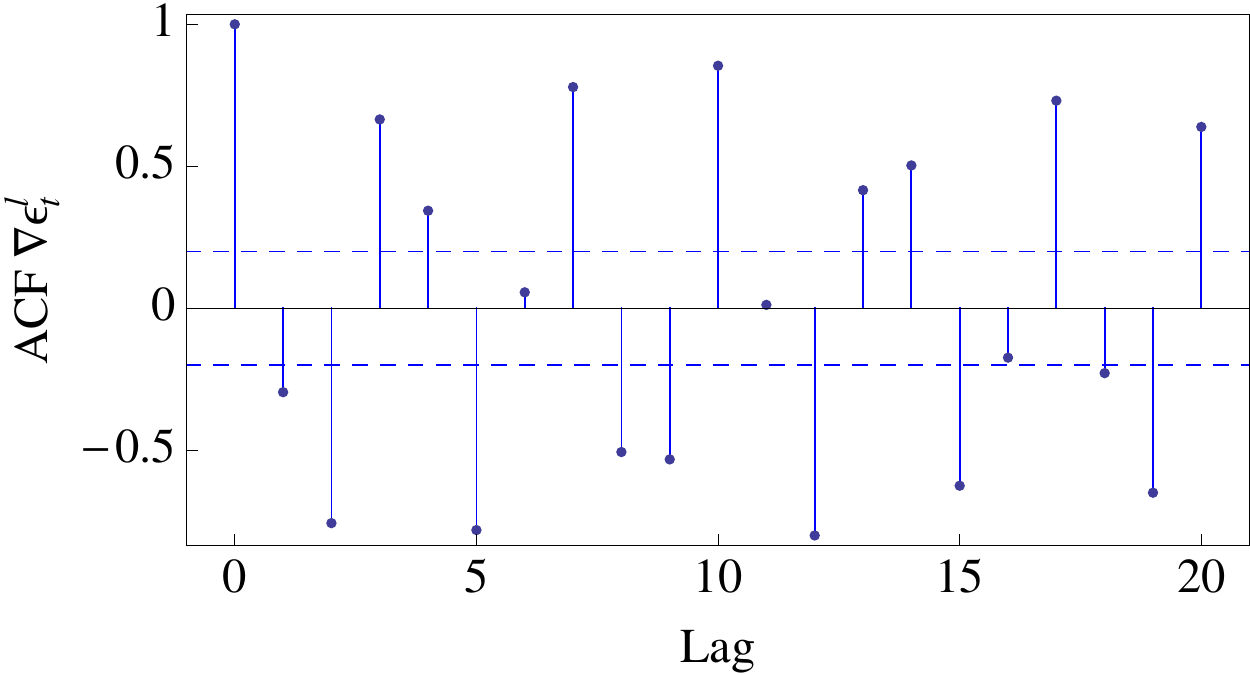}\\

\includegraphics[scale=.5]{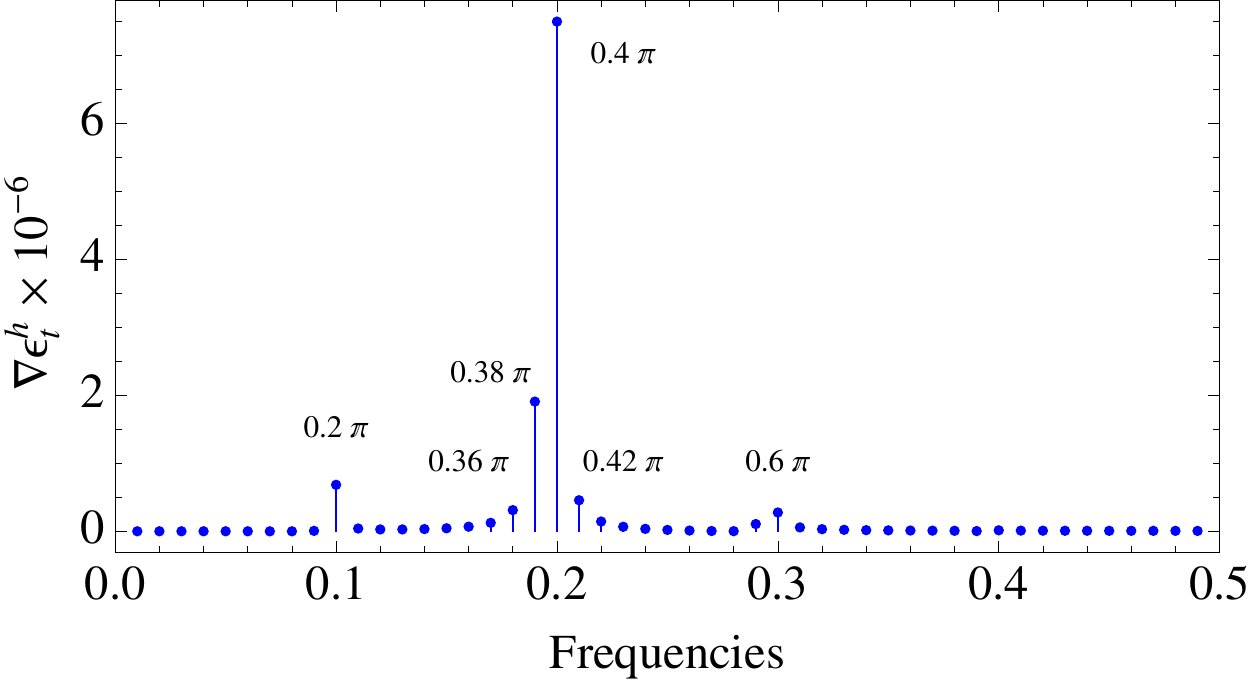}
\includegraphics[scale=.5]{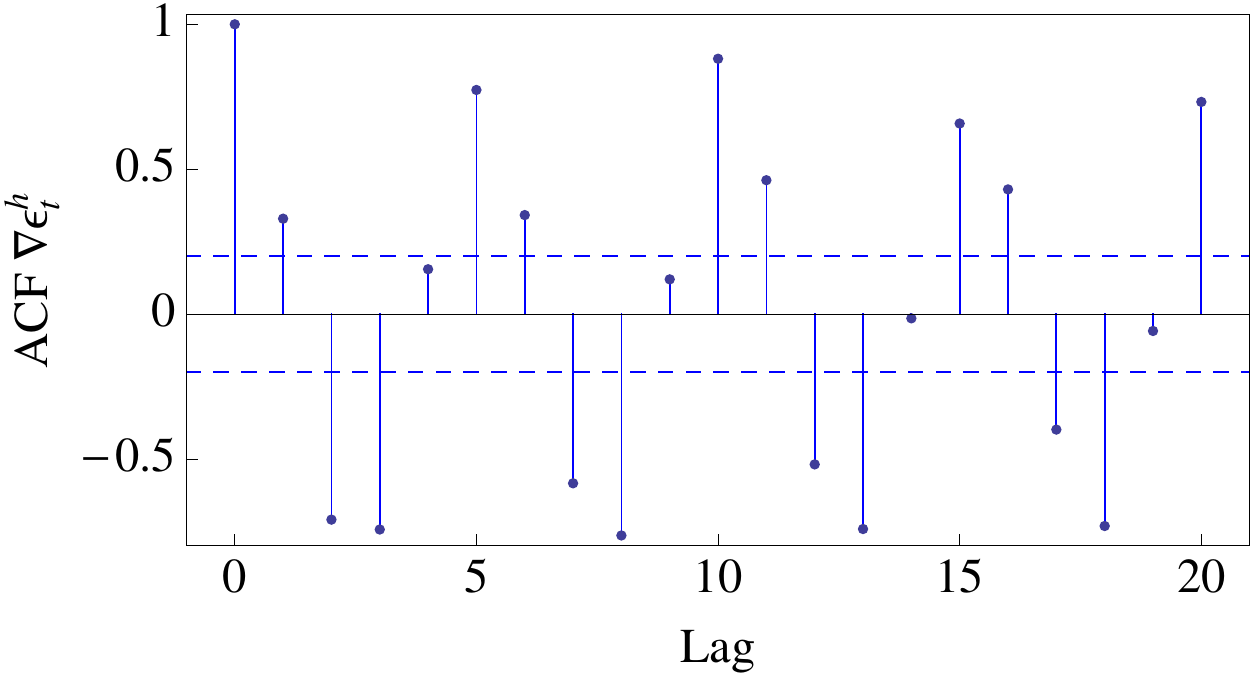}\\

\includegraphics[scale=.5]{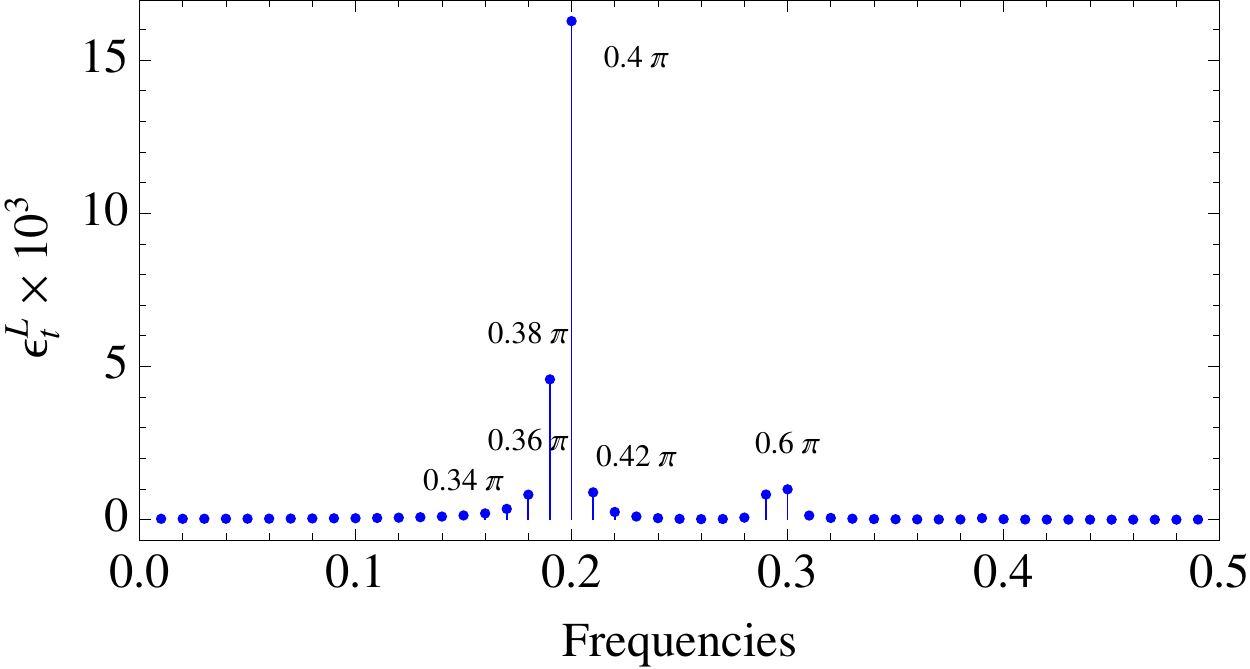}
\includegraphics[scale=.5]{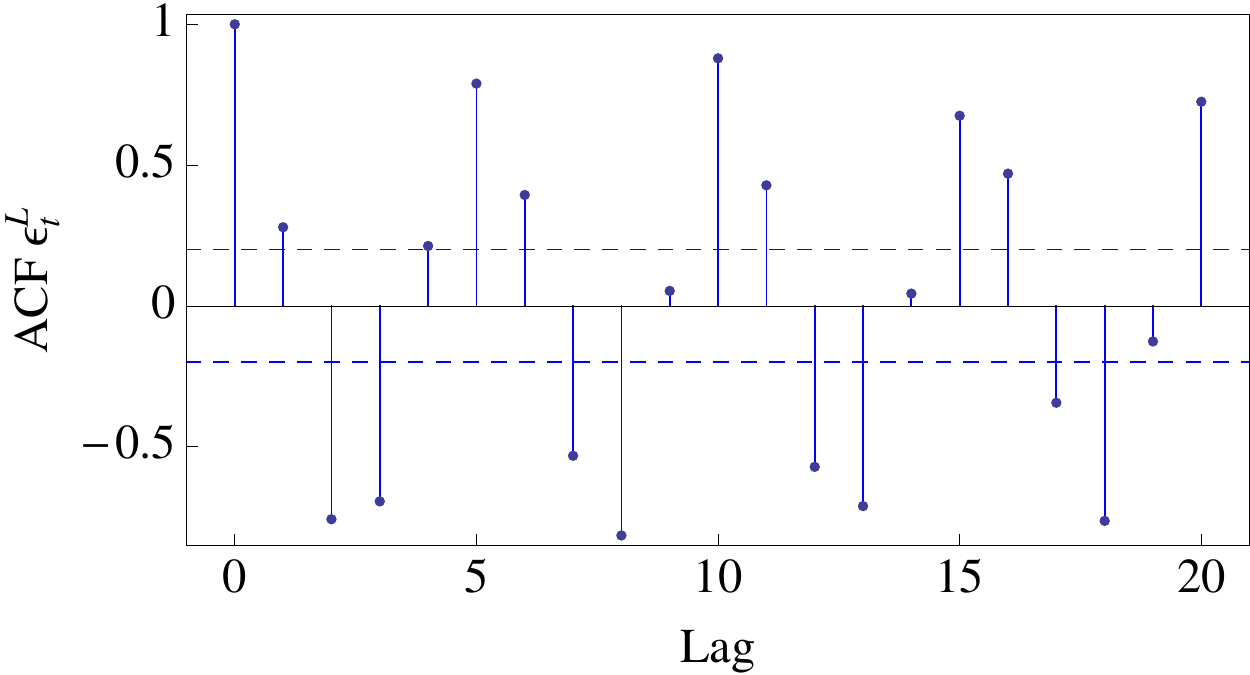}
\caption{Periodograms (left) and autocorrelation functions (right)  of the series $\varepsilon_t^l$, $\varepsilon_t^h$ and $\varepsilon_t^L$.  }\label{figure2b}
\end{figure}

The first step in the preliminary analysis consists in plotting the time series $\varepsilon_t^l$, $\varepsilon_t^g$, $\varepsilon_t^h$, $\varepsilon_t^L$ and 
$\varepsilon_t^G$. Initially, values have been generated every 10 minutes.
Figures \ref{figure2a} and \ref{figure2b}  only show the sequence plots, periodograms and autocorrelation functions  of the series $\varepsilon_t^l$, $\varepsilon_t^h$ and $\varepsilon_t^L$, because the behaviour  of    $\varepsilon_t^g$ and  $\varepsilon_t^G$  is approximately the same as $\varepsilon_t^l$ and $\varepsilon_t^L$ respectively.

First,  we begin analysing  the sequence plots. The series $\varepsilon_t^l$ has a  linear increasing trend and a periodic behaviour with a period of approximately  $33.33$ minutes. On the other hand, the trend of $\varepsilon_t^h$ is linear and decreasing, but its periodic behaviour is difficult to discover because it is hidden by its trend, so other statistical tools  are needed for its detection. Finally,  the  trend of $\varepsilon_t^L$ is not significant, so it can be considered  constant; however, this series shows seasonal fluctuations, the first with a period of almost   50 minutes, whereas  a repetitive pattern of alternate oscillations can   be clearly observed    approximately every 100 minutes.
 
Then, the periodograms and the autocorrelation functions are analysed. It is worth noting that these functions can only be applied to stationary series so, in the first place, it will be necessary to differentiate the series $\varepsilon^l_t$ and $\varepsilon^h_t$ so as to remove their trends.

The periodogram of $\nabla \varepsilon^l_t$ (Figure \ref{figure2b}) shows that the most significant frequency is close to  $0.6 \pi$, that is, a periodic behaviour  with an approximate period of $33.33$  minutes, whereas series  $\nabla \varepsilon^h_t$ and  $\varepsilon^L_t$  have their main frequency near $0.4 \pi$, which corresponds to a period of almost  $50$ minutes.

Finally, from the analysis of the autocorrelation functions of  series $\nabla \varepsilon^l_t$, $\nabla \varepsilon^h_t$ and  $\varepsilon^L_t$, it can be observed that, despite having high correlation in several delays,  the strongest one corresponds to lag 10, which implies  a close relationship each 10 points (approximately each 100 minutes).

This preliminary analysis allows us to conclude that, although, in principle, it might seem that there are three main periodicities of approximate periods   $ 33.33 $, $50$ and $100$ minutes,  in reality, the most remarkable is the last one, which corresponds approximately to the Keplerian period of  satellite 1, $101.926$ minutes.

Then, after estimating the initial values $A_0,\,B_0,\,S_{-s+1},\ldots, S_{-s}$ and $S_{0}$ by means of the heuristic method described in Section 4, the next step consists in the identification of the  optimal values for the smoothing parameters $\alpha$, $\beta$ and $\gamma$ of the Holt-Winters method that minimize the distance error. As shown in Algorithm  1,  their values are obtained   by applying  the L-BFGS-B algorithm to one of the error functions (\ref{med.err}), MSE in this case.

Finally,  this model  is analysed in order to experimentally determine  the best  amount of  control data by choosing the number of  both satellite revolutions and  points per revolution  which minimize the distance error over 30 days,  taking into account that we are considering here  the Keplerian period as the revolution period. We proceed by fixing the number of revolutions in the first place, with the aim of determining the optimal amount of points per revolution. Once it has been done, the best number of revolutions can be determined. Several configurations have been tested, keeping in mind as general guidelines that the control period has to be long enough so as to contain any dynamics to be modelled, and the sampling rate needs to be high enough so as to capture the highest frequency that the hybrid propagator is designed to model and reproduce. This analysis leads us to consider 10 satellite revolutions and 12 points per revolution as the best choice to constitute the control data for all the studied satellites, which represents a time span of approximately 17 hours in the case of satellite 1. More details about this analysis can be found in \cite{san2014gru_montse_phd}, although further research on this issue is being conducted in order to draw general conclusions regarding the optimal configuration of the control period.

Table \ref{table3} shows the distance error between HPPD0 and the numerical integration of the main problem at different instants. As can be seen, the distance error, even after a 30-day propagation, is lower than the pure analytical propagator PPD0 distance error only after 17 hours, i.e. the control period.

\begin{table}[htbp!!]
\caption{ Maximum distance  error between HPPD0 and the numerical integration of the main problem for  satellite 1 over different propagation spans.}\label{table3}
\begin{center}
\begin{tabular}{cc}
Time &   \multicolumn{1}{c}{Distance error (km) }\\ \hline%\\[-2ex]
$17$ hours & $\:\:2.69$  \\
  $\:1$ day & $\:\:2.85$  \\
 $\:2$ days & $\:\:3.10$ \\
 $\:7$ days & $10.83$ \\
 $30$ days & $13.79$ 
 \end{tabular}
\end{center}
\label{default}
\end{table}

\begin{figure}[!!htp]
\centering
\includegraphics[scale=.5]{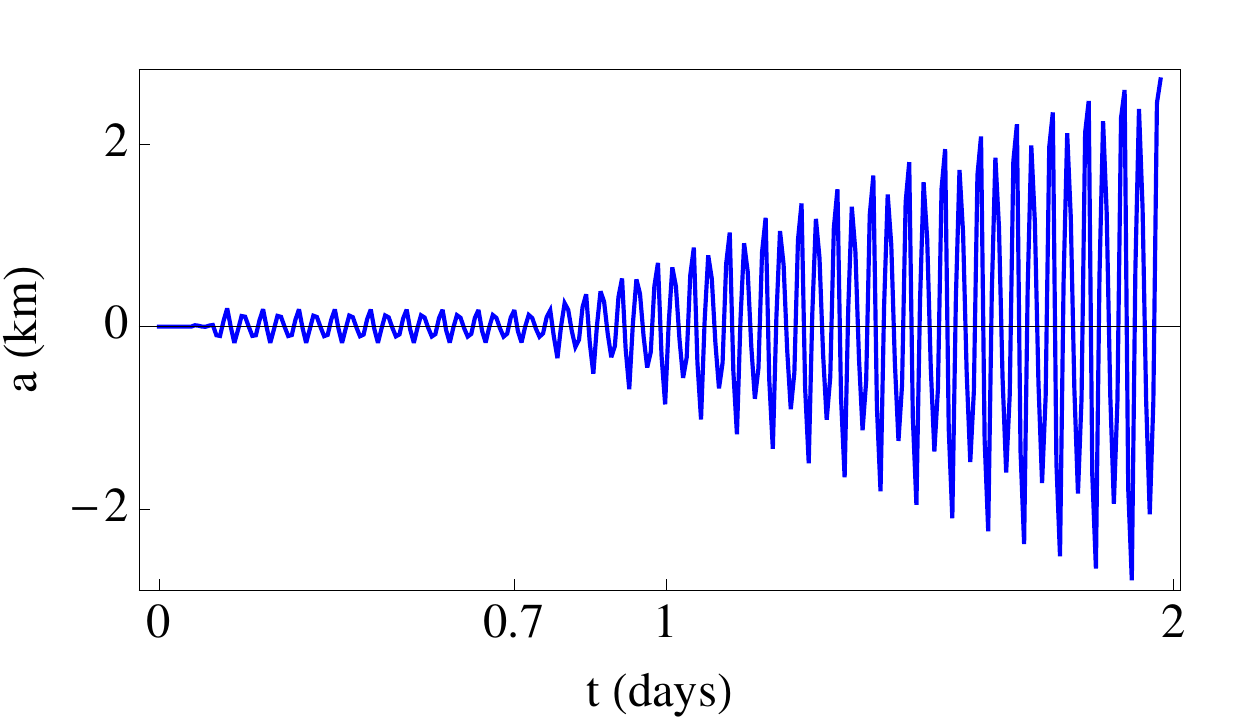}
\includegraphics[scale=.5]{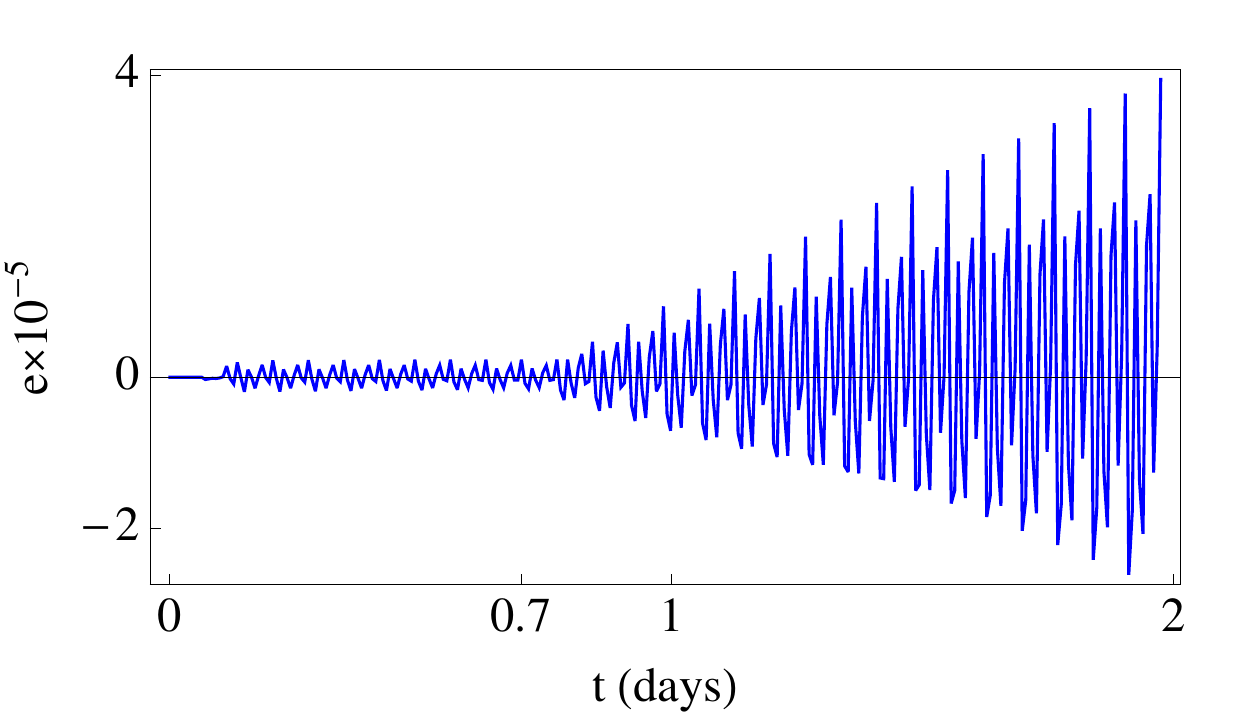}\\

\includegraphics[scale=.5]{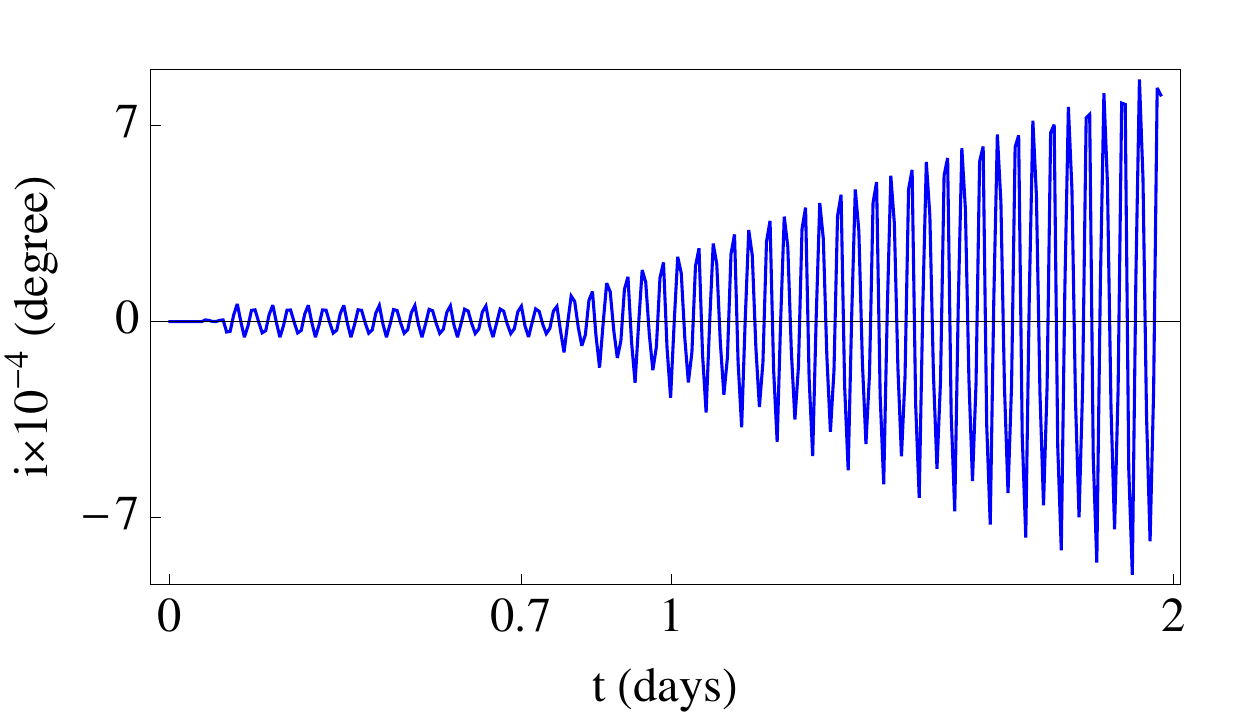}
\includegraphics[scale=.5]{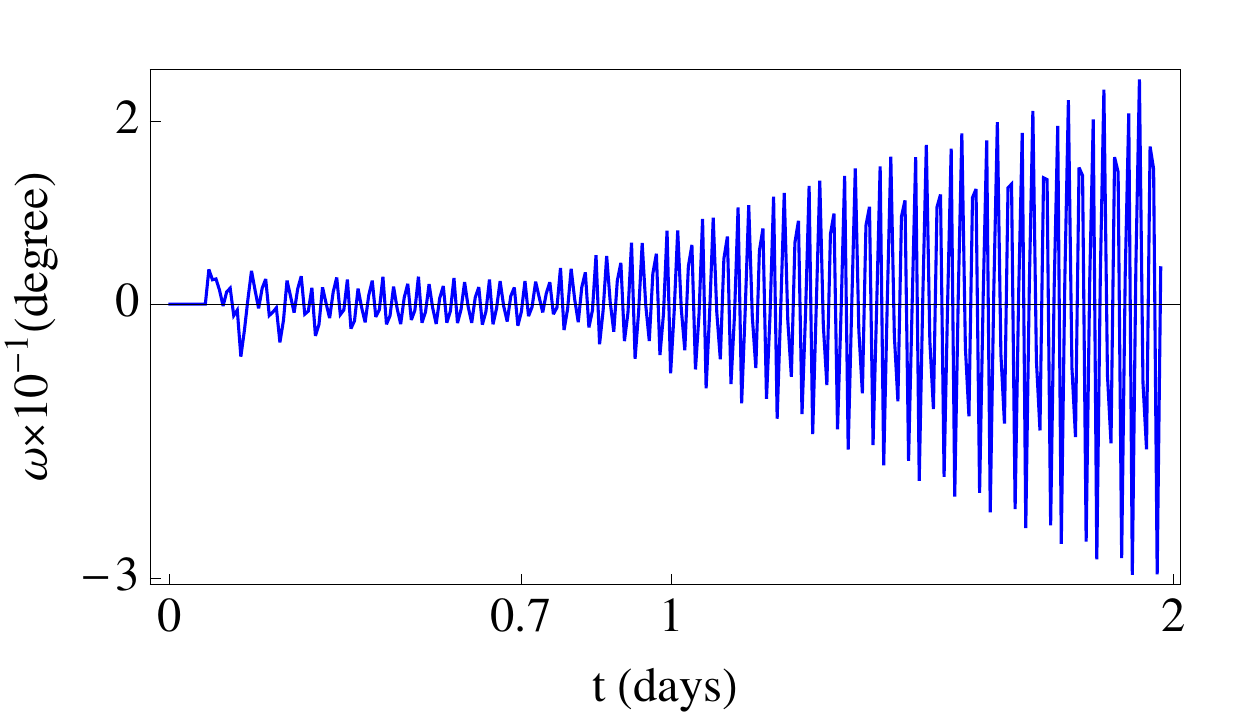}\\

\includegraphics[scale=.5]{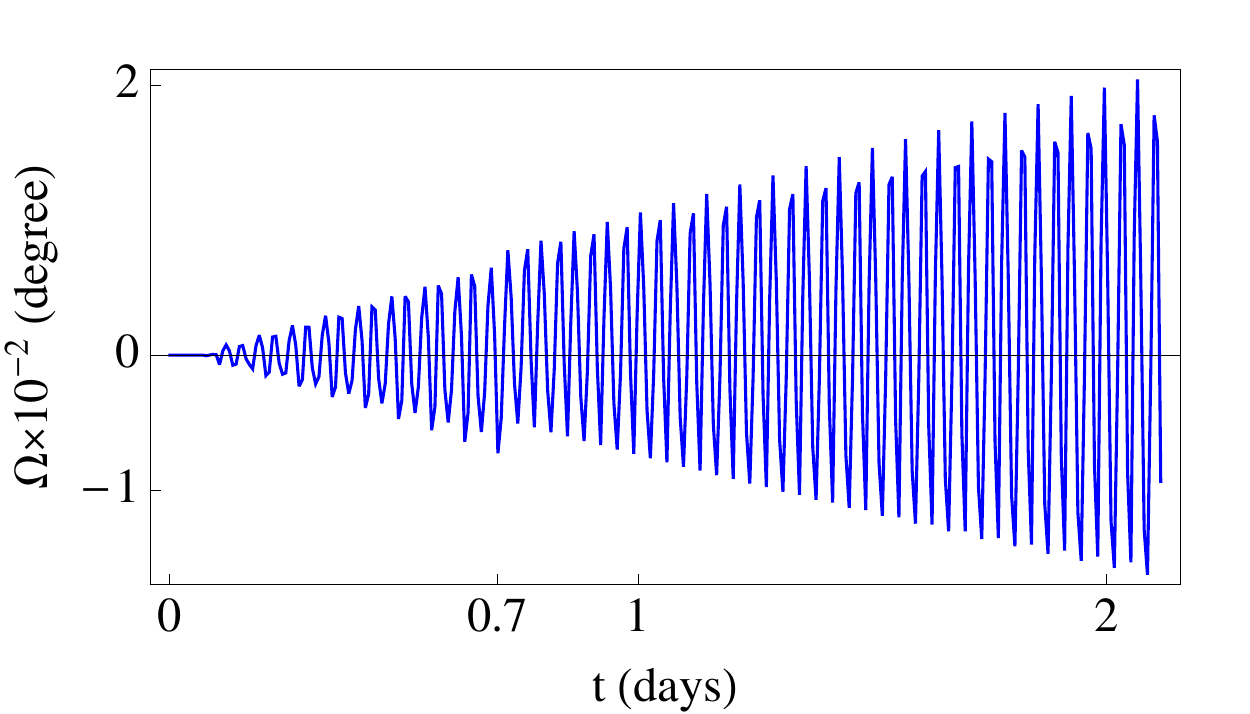}
\includegraphics[scale=.5]{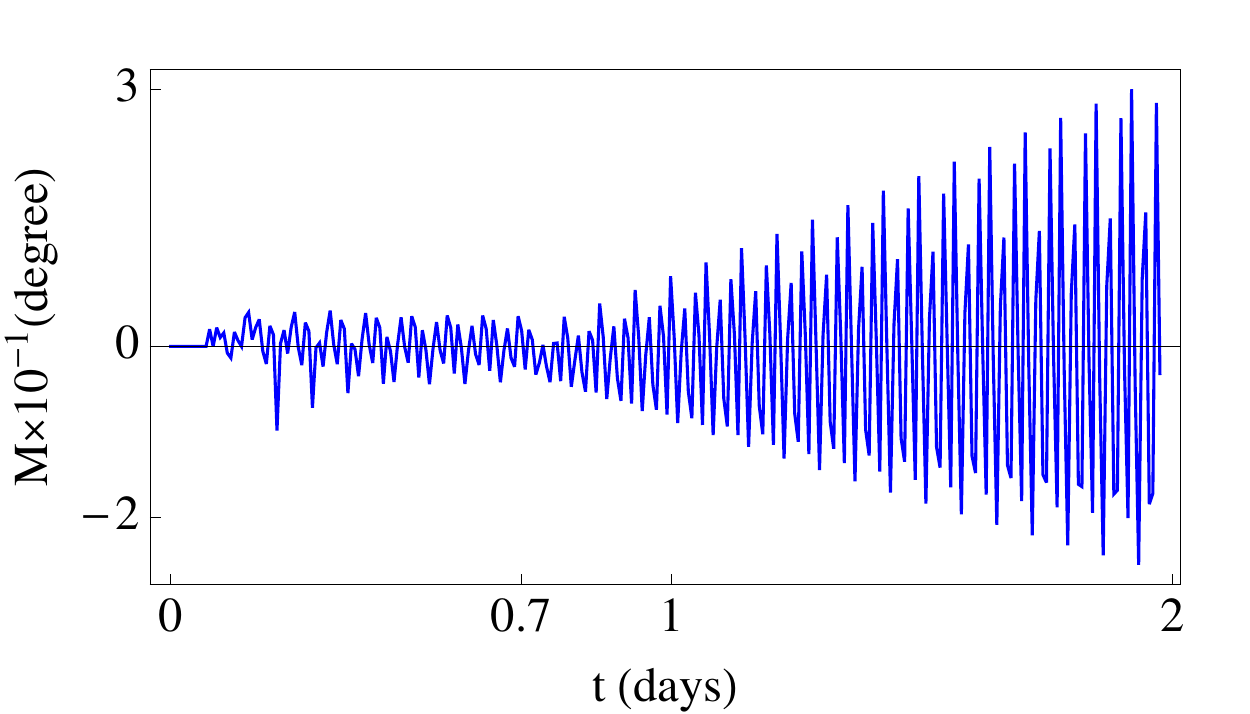}

\caption{Evolution of  the orbital-element differences between HPPD0 and the numerical integration of the main problem for satellite 1.}\label{figure4}
\end{figure}

Figure \ref{figure4} shows how the differences between the orbital elements of both HPPD0 and the numerical integration of the main problem evolve during 2 days for satellite 1.

\begin{figure}[!!htp]
\centering
\includegraphics[scale=.5]{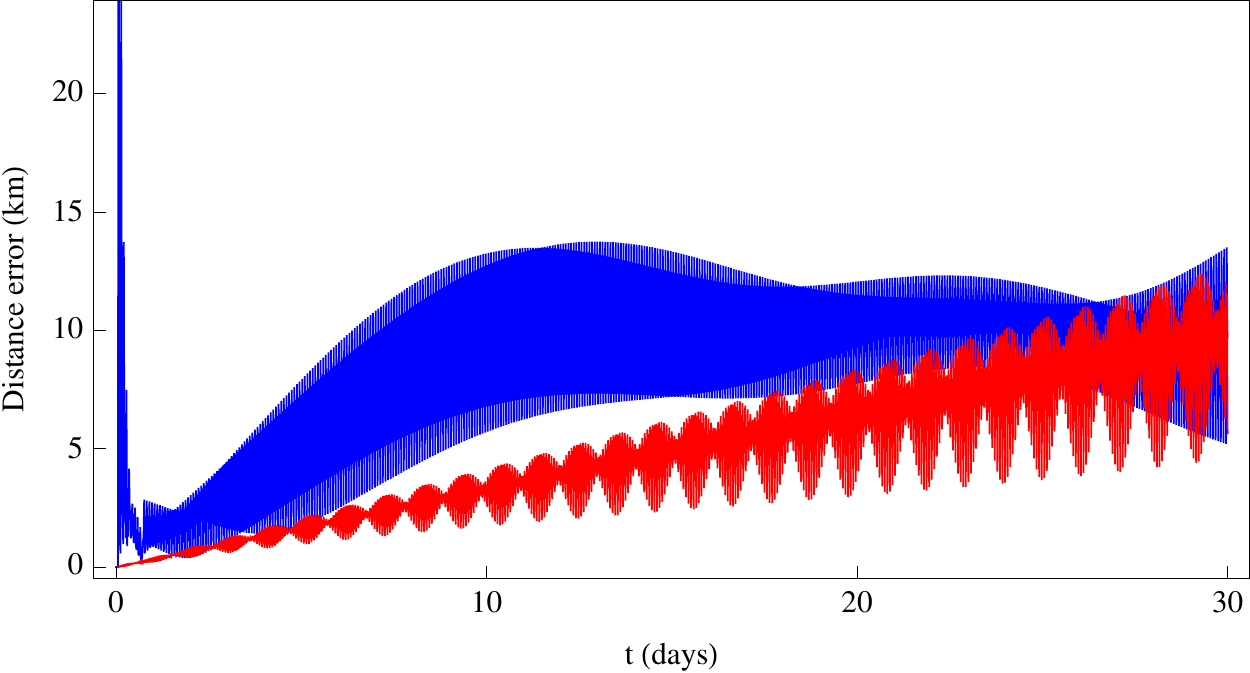}
\caption{Distance error of the zero-order hybrid propagator HPPD0 (blue line) and the first-order pure analytical propagator PPD1 (red line) for satellite 1.}\label{fig01}
\end{figure}

Finally, Figure \ref{fig01} compares the distance error for both the zero-order hybrid propagator HPPD0 and the first-order pure analytical propagator PPD1. The error is lower for the analytical propagator during the first 25 days, as could be expected from a higher-order approximation, although both errors become similar as the last part of the 30-day studied period is reached.

\subsection{Improving the analytical approximation}

In this subsection we study the hybrid propagators based on the first-order and second-order analytical propagators, PPD1 and PPD2 respectively. All the analysed hybrid propagators improve the accuracy of the analytical theory they are based on; nevertheless, preliminary analysis leads us to conclude that, similarly to the zero-order propagator, optimum results are achieved when the exact revolution period is taken into account, in such a way that the data sampling rate  is chosen in order to allow for a complete number of samples per satellite revolution.

Furthermore, the fact that not only periodograms and autocorrelation functions, but also the smoothing parameters to be obtained, are quite similar for the three studied orders of approximation, indicates that the error series $\varepsilon_t^l$, $\varepsilon_t^g$, $\varepsilon_t^h$, $\varepsilon_t^L$ and  $\varepsilon_t^G$ maintain the same characteristics to be estimated, albeit with decreasing magnitudes for higher-order models, and consequently must be processed and forecasted by means of similar models.

According to what has been expounded, hybrid propagators that consider $MSE$ as the error function to be minimized, and take exactly 12 samples, at a regular rate, for each of the 10 revolutions that constitute the control period, are again the most accurate, leading to the best results for short, medium and long-term estimation.

\begin{figure}[!!htp]
\centering
\includegraphics[scale=.5]{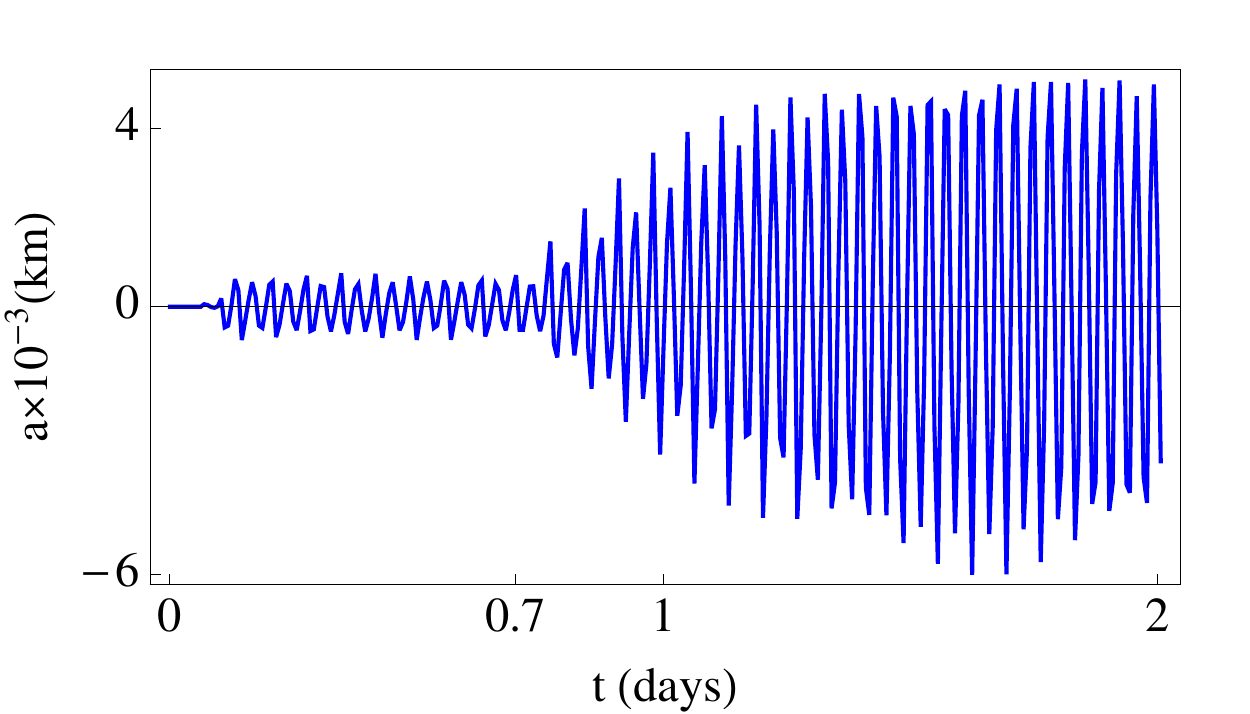}
\includegraphics[scale=.5]{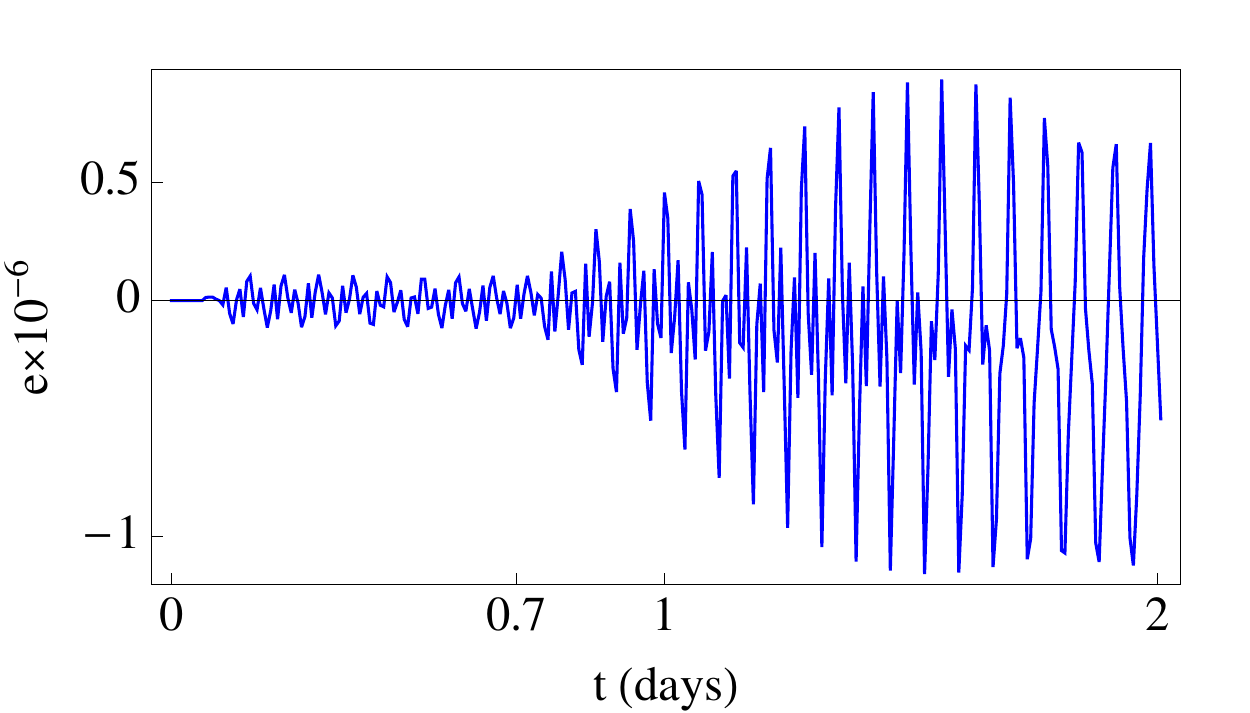}\\

\includegraphics[scale=.5]{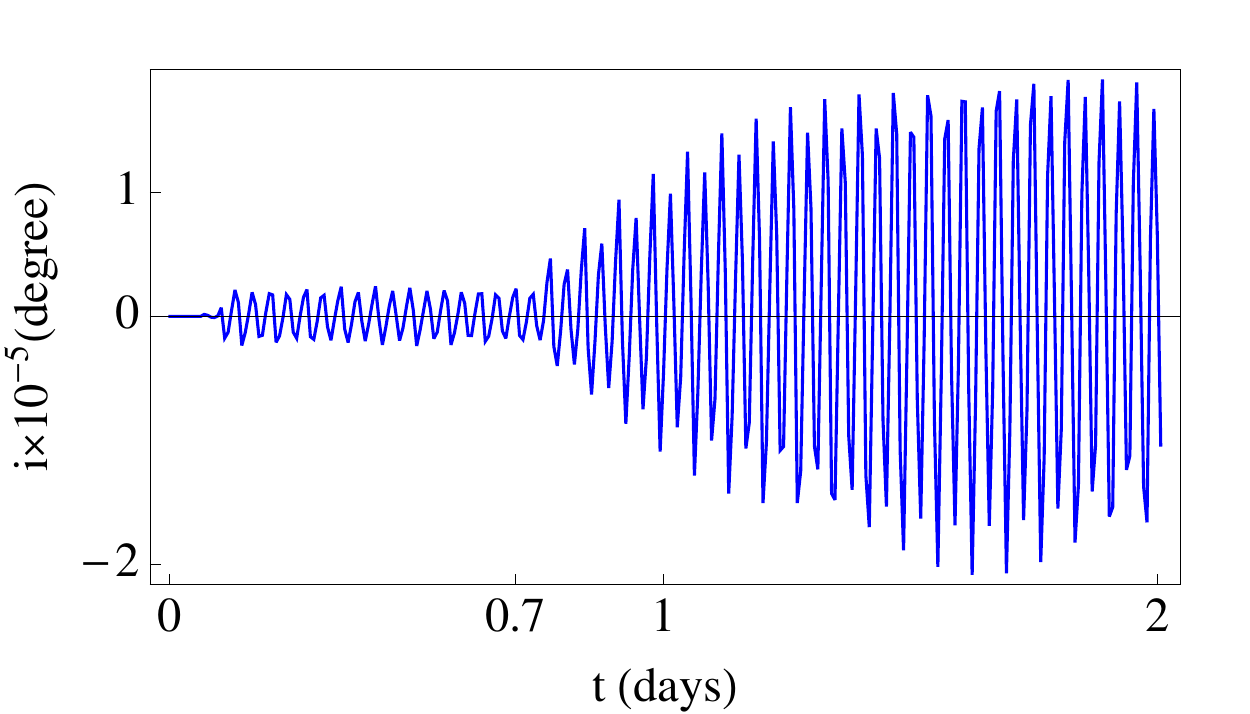}
\includegraphics[scale=.5]{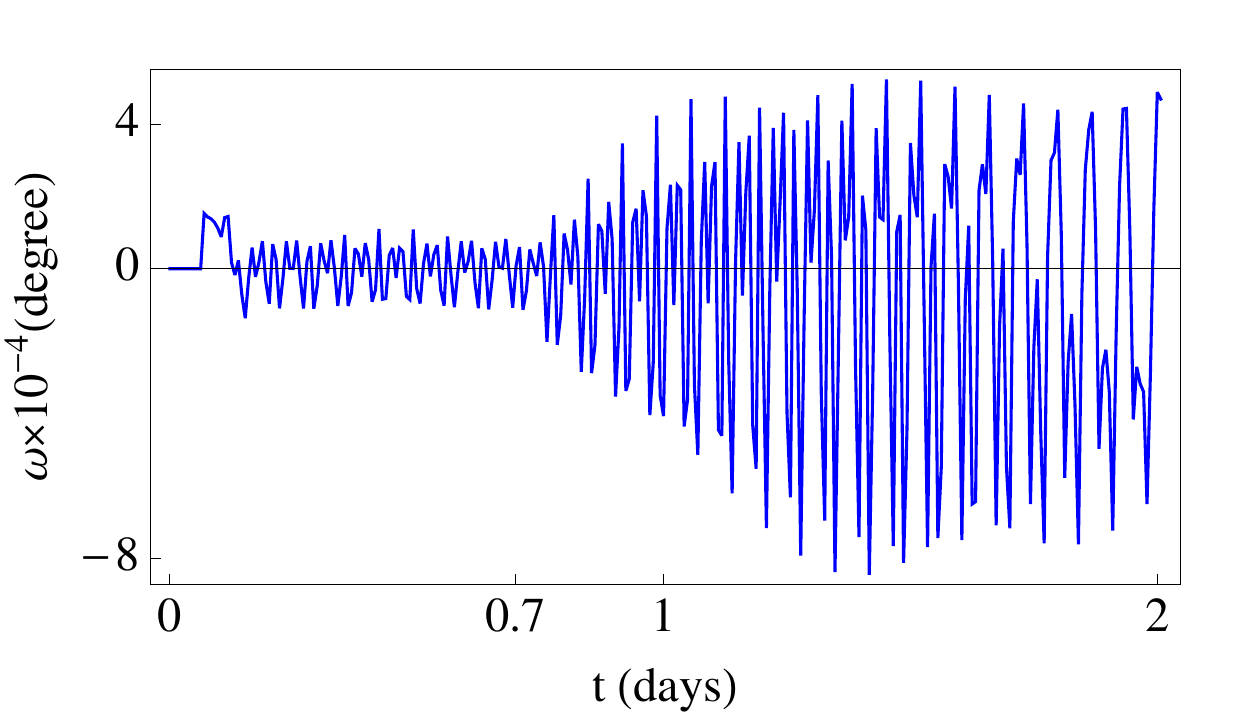}\\

\includegraphics[scale=.5]{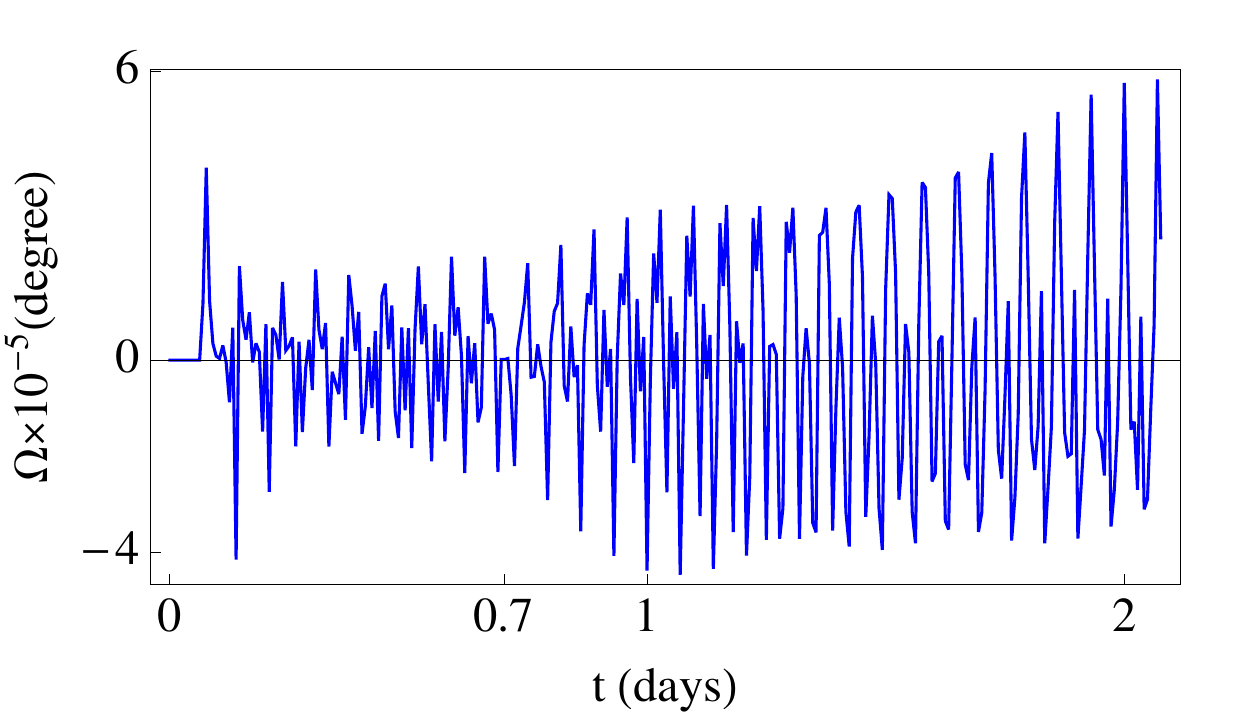}
\includegraphics[scale=.5]{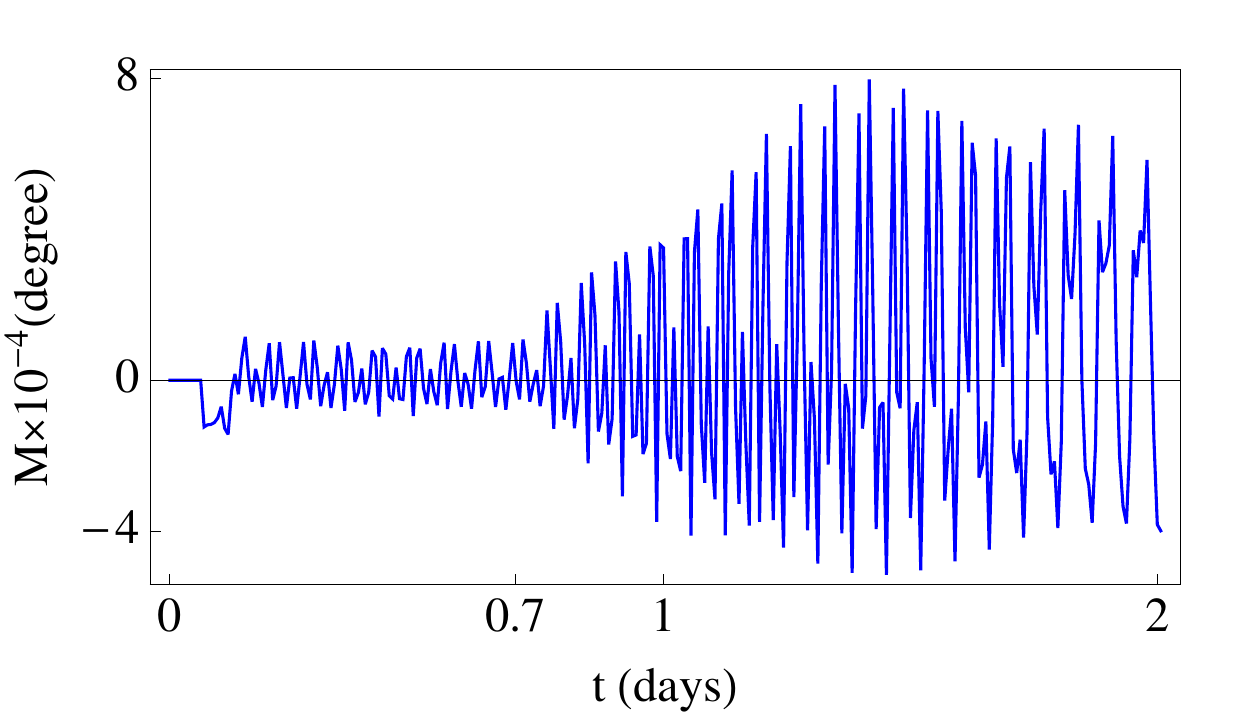}

\caption{Evolution of  the orbital-element differences between HPPD1 and the numerical integration of the main problem for  satellite  1.}\label{fig010}
\end{figure}

Figure \ref{fig010} plots the differences in the evolution of the orbital elements of both the first-order hybrid propagator HPPD1 and the accurate numerical integration of the main problem for satellite 1.

Table \ref{errorANA010} shows the predictive capability of both the analytical and associated optimum hybrid propagators. It is worth noting that, after a 30-day propagation, distance error can be reduced by a factor of 20 in the case of first-order theory, or nearly 70 for second-order theory. 

\begin{table}[htbp!!]
\caption{Distance error of the pure analytical and the hybrid propagators  for both first  and second-order theories (satellite 1).}\label{errorANA010}
\begin{center}
\begin{tabular}{crrrr}
Time &     \multicolumn{1}{c}{PPD1(km) }&     \multicolumn{1}{c}{HPPD1(km) }&     
\multicolumn{1}{c}{PPD2(km) }&     \multicolumn{1}{c}{HPPD2(km) }\\ \hline
$17$ hours & $0.2758$ & $0.0008$ & $0.0007$ & $4.5\times 10^{-6}$  \\
  $1$ day &  $0.4037$ & $0.0015$   & $0.0010$ & $8.1\times 10^{-6}$ \\
 $2$ days &  $0.8223$ & $0.0061$  & $0.0020$ & $1.9\times 10^{-5}$\\
 $7$ days &  $2.9175$ & $0.0548$ & $0.0070$ & $4.2\times 10^{-5}$\\
 $30$ days &  $12.5706$ & $0.6462$ & $0.0290$ & $4.2\times 10^{-4}$
 \end{tabular}
\end{center}
\label{default}
\end{table}

\begin{figure}[!!htp]
\centering
\includegraphics[scale=.5]{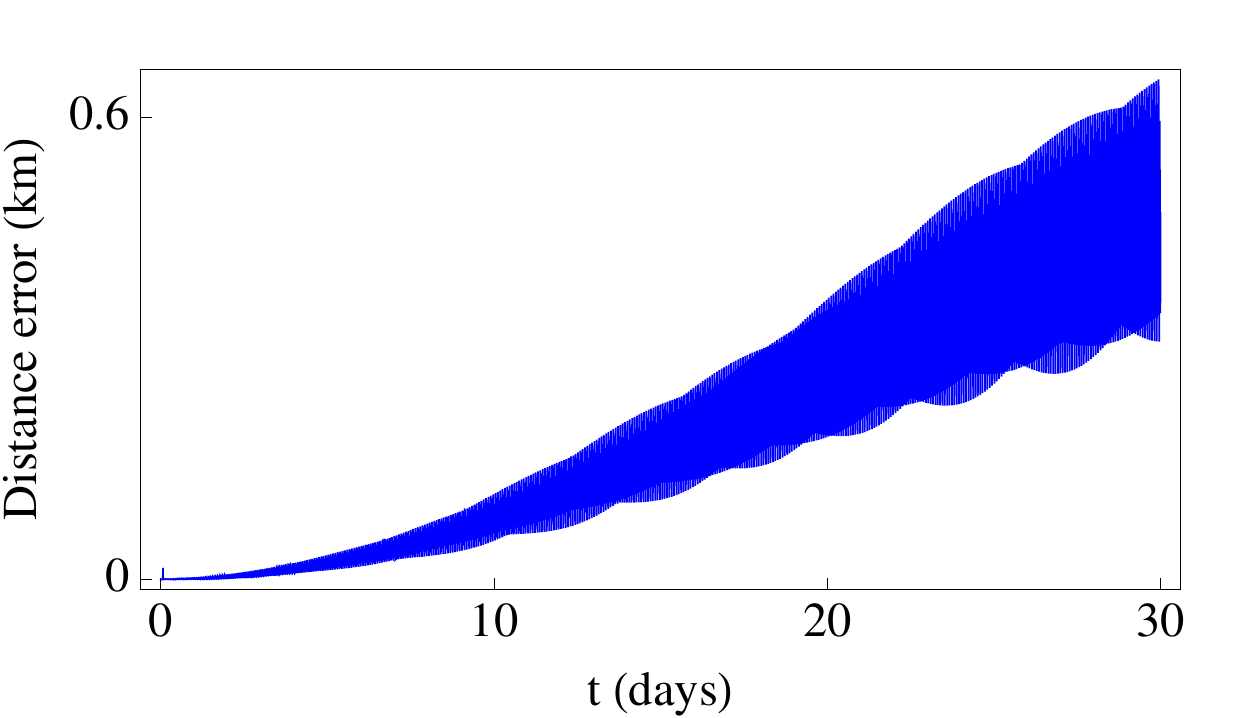}
\caption{Distance error of the first-order hybrid propagator HPPD1 for satellite 1}\label{fig102}
\end{figure}

Figure \ref{fig102}  plots the evolution of the distance error for the first-order hybrid propagator HPPD1 during a 30-day time span. It should be noted that the error remains quite low, just about 10 metres, for more than 5 days.

\subsection{Hybrid propagators for the complete set of satellites}

The importance of acquiring the seasonal variations originated by the satellite revolution period has been verified for satellite 1 in previous subsections. Therefore, for the rest of the satellites, data will also be considered with sampling rates adapted to each satellite period, in such a way that a complete number of samples per revolution is always guaranteed. Apart from that, sequence graphics, periodograms and autocorrelation functions show that, in addition to the revolution-time periodicity, other two periodicities with magnitudes half and third the same satellite revolution time also exist. For those reasons, and with the aim of acquiring such periodic behaviour, it would be desirable to choose a number of samples per revolution which is multiple of both 2 and 3. Taking into account that it is never advisable to consider an amount of data either too high or too low, it is concluded that 12 samples per satellite revolution is an appropriate value as sampling rate  also for the remaining 8 satellites.

\begin{figure}[!!htp]
\centering
\includegraphics[scale=.61]{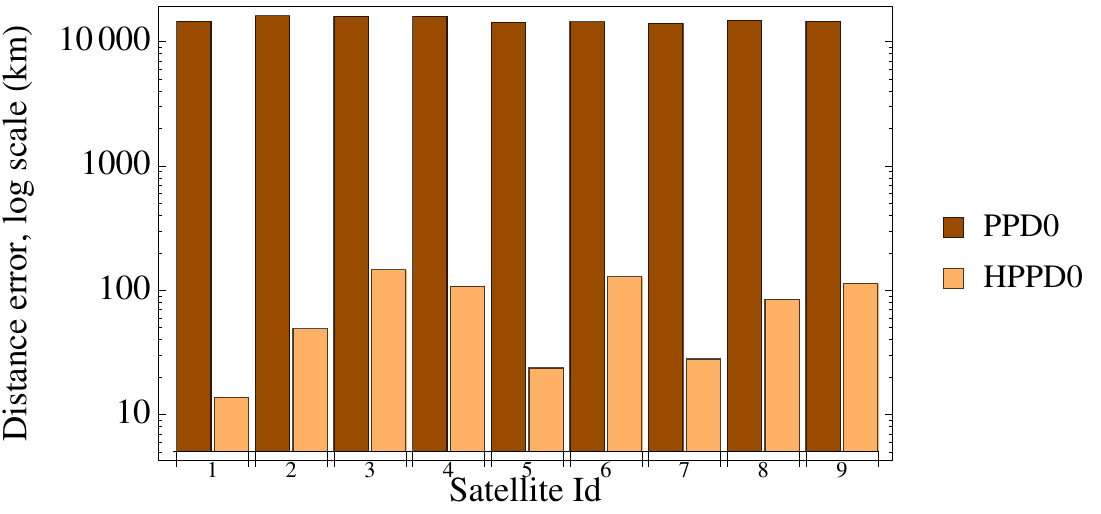}
\includegraphics[scale=.59]{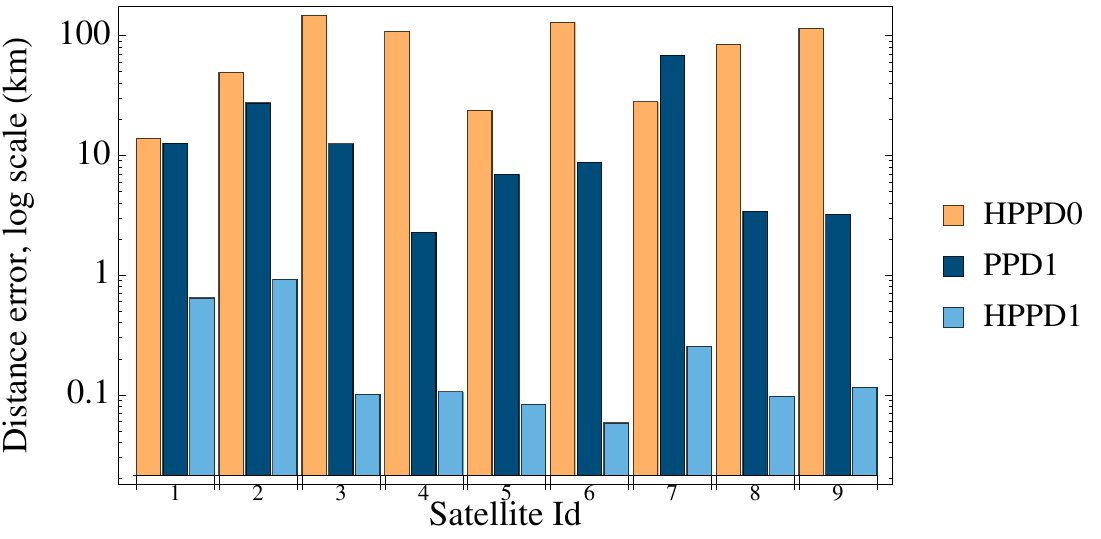}

\caption{Distance error of the pure analytical and the hybrid propagators for both zero and first-order theories after a 30-day propagation (satellites 1-9). }\label{figlast}
\end{figure}

Figure \ref{figlast} and Table \ref{tab:kepa23} show the distance error of the pure analytical and the hybrid propagators for both zero and first-order theories after a 30-day propagation for the complete set of satellites analysed in this work. Taking into account that Figure \ref{figlast} has been plotted with logarithmic scale, it should be noted that, in general, the distance error is between 2 and 3 orders of magnitude lower for the hybrid propagator than for the pure analytical propagator in the case of the zero-order theory, i.e. HPPD0 versus PPD0. This hybrid propagator HPPD0 has distance errors between only 0 and 1 orders of magnitude higher than the first-order analytical propagator PPD1, which is one order of approximation higher, although in the case of satellite 7 the hybrid propagator is even more accurate than the superior analytical propagator. When the two first-order propagators are compared, it is found that the hybrid HPPD1 is between 1 and 2 orders of magnitude more accurate than the analytical PPD1.

\begin{table}[htbp!!]
\caption{Distance error of the pure analytical and the hybrid propagators for both zero and first-order theories after a 30-day propagation (satellites 1-9).}\label{tab:kepa23}
\begin{center}
\begin{tabular}{ccccccc}
Id &   PPD0 (km)   &  HPPD0 (km)   & PPD1(km)  &  HPPD1 (km) \\ \hline\\[-1.5ex]
1 & $14506.5$ & $13.792$ & $12.6$  & $0.634$\\
2  &$16183.7$& $49.136$ & $27.3$&  $0.918$\\
3  &$15987.4$&$146.465$ & $12.5$  & $0.101$ \\
4 &$15922.4$& $107.905$ & $2.3$ &  $0.106$\\
5 & $14292.2$& $23.774$ & $6.9$  &  $0.083$\\
6 & $14456.9$&$128.633$ & $8.7$  & $0.058$ \\
7 & $14012.8$& $27.992$ &  $68.6$ &$0.255$\\
8 & $14882.6$& $84.369$ & $3.4$ & $0.096$\\
9 & $14489.9$& $114.199$ &$3.2$  &  $0.115$ 
\end{tabular}
\end{center}
\end{table}

\section{Conclusion}

In this work, we have presented a new approach called hybrid perturbation theory. The proposed methodology, which combines an integration method and a prediction technique, has been illustrated through the combination of a simplified general perturbation theory and a statistical time series model based on an additive Holt-Winters method. The hybrid propagators that have been developed have proven an increase in the accuracy of the analytical theory for predicting the position and velocity of the studied orbiters, as well as modelling higher-order terms and other external forces not considered in the analytical theory.

It has been found that the effect of considering a complete number of samples per revolution in hybrid propagators varies depending on the order of the underlying analytical theory, and thus on its margin for improvement. In the case of hybrid propagators based on the zero-order analytical theory, a dramatic reduction in distance error is achieved. In contrast, the second-order hybrid propagator, whose margin for improvement is very reduced, only reaches a slight increase in accuracy when the sample rate is such that a complete number of values fits into a satellite revolution.

Another remarkable conclusion is that similar smoothing parameters are obtained for hybrid propagators based on a certain analytical theory with different orders of approximation, which implies that the error time series to be modelled maintain the same characteristics, even though their magnitude can vary.

\section*{Acknowledgments}

This work has been funded by the Spanish Finance and Competitiveness Ministry under Project ESP2014-57071-R. The authors would like to thank the  reviewers for their valuable suggestions.

%\section*{References}

\bibliographystyle{elsarticle-harv}
\bibliography{references}

\end{document}